\documentclass{aa}
\usepackage{graphicx}
\usepackage{bm}
\usepackage{times}
\usepackage{amsmath}
\usepackage{amssymb}
\usepackage{mathrsfs}
\usepackage{natbib}
\usepackage{color}
\usepackage[colorlinks,allcolors=blue]{hyperref}
\bibpunct{(}{)}{;}{a}{}{,}
\graphicspath{{./fig/}{./png/}}
\usepackage{hyperref}
\newcommand{\eee}{{\bm e}}

\newcommand{\fff}{{\bm f}}

\newcommand{\kkk}{{\bm k}}

\newcommand{\uuu}{{\bm u}}

\newcommand{\xxx}{{\bm x}}
\newcommand{\aaaT}{{\bm a}^{\rm T}}
\newcommand{\bbbT}{{\bm b}^{\rm T}}

\newcommand{\BBB}{{\bm B}}
\newcommand{\mBBB}{\overline{\bm B}}
\newcommand{\BBBT}{{\bm B}^{\rm T}}
\newcommand{\mBBBT}{\overline{\bm B}^{\rm T}}

\newcommand{\mFFF}{\overline{\bm F}}

\newcommand{\JJJ}{{\bm J}}

\newcommand{\mUUU}{\overline{\bm U}}

\newcommand{\mH}{\overline{H}}

\newcommand{\mBBj}{\overline{B}_j}
\newcommand{\mBBk}{\overline{B}_k}

\newcommand{\mJJj}{\overline{J}_j}

\newcommand{\mEMF}{\overline{\bm{\mathcal{E}}}}
\newcommand{\mEE}{\overline{\mathcal{E}}}
\newcommand{\EMFi}{\overline{\mathcal{E}}_i}
\newcommand{\alij}{\alpha_{ij}}
\newcommand{\etij}{\eta_{ij}}
\newcommand{\beijk}{\beta_{ijk}}
\newcommand{\zVj}{z^W_j}
\newcommand{\zUj}{z^U_j}
\newcommand{\Eq}[1]{Eq.~(\ref{#1})}

\newcommand{\Eqsa}[2]{Eqs.~(\ref{#1}) and (\ref{#2})} 

\newcommand{\EQ}{\begin{equation}}
\newcommand{\EN}{\end{equation}}
\newcommand{\EQA}{\begin{eqnarray}}
\newcommand{\ENA}{\end{eqnarray}}

\newcommand{\pd}{\partial}

\newcommand{\mean}[1]{\overline{#1}}
\newcommand{\meanv}[1]{\overline{\bm #1}}

\newcommand{\cs}{c_{\rm s}}

\newcommand{\nut}{\nu_{\rm t}}
\newcommand{\nutz}{\nu_{\rm t0}}
\newcommand{\tnut}{\tilde{\nu}_{\rm t}}
\newcommand{\etat}{\eta_{\rm t}}
\newcommand{\etatz}{\eta_{\rm t0}}

\newcommand{\tetat}{\tilde{\eta}_{\rm t}}
\newcommand{\urms}{u_{\rm rms}}

\newcommand{\Ma}{{\rm Ma}}

\newcommand{\kf}{k_{\rm f}}
\newcommand{\tkf}{\tilde{k}_{\rm f}}

\newcommand{\tkU}{\tilde{k}_U}
\newcommand{\tkB}{\tilde{k}_B}

\newcommand{\Pm}{{\rm Pm}}
\newcommand{\Pmt}{{\rm Pm}_{\rm t}}

\newcommand{\Rey}{{\rm Re}}

\newcommand{\Rm}{{\rm Rm}}

\newcommand{\Sh}{{\rm Sh}}

\newcommand{\qyz}{Q_{yz}}

\newcommand{\css}{c_{\rm s}^2}

\newcommand{\nab}{\mbox{\boldmath $\nabla$} {}}

{}

\newcommand{\meanH}{\mean{H}}
\def\onethird{{\textstyle{1\over3}}}

\def\onehalf{{\textstyle{1\over2}}}

\newcommand{\Fig}[1]{Figure~\ref{#1}} 

\newcommand{\Seca}[1]{Sect.~\ref{#1}}
\newcommand{\Sec}[1]{Section~\ref{#1}} 
\newcommand{\Table}[1]{Table~\ref{#1}}

\newcommand{\s}{\,{\rm s}}
\newcommand{\m}{\,{\rm m}}

\newcommand{\uu}{{\bm u}}

\newcommand{\mlnrho}{\overline{\ln\rho}}
\renewcommand{\mFFF}{\boldsymbol{\cal F}}

\begin{document}


\authorrunning{K\"apyl\"a et al.}
\titlerunning{Turbulent viscosity from simulations}

  \title{Turbulent viscosity and 
   magnetic Prandtl number from simulations of isotropically forced turbulence}

   \author{P. J. K\"apyl\"a
          \inst{1,2}
          \and
          M. Rheinhardt
          \inst{2}
          \and
          A. Brandenburg
          \inst{3,4,5,6}
          \and
          M. J. K\"apyl\"a
          \inst{7,2}
          }
   \institute{Georg-August-Universit\"at G\"ottingen, Institut f\"ur
  Astrophysik, Friedrich-Hund-Platz 1, D-37077 G\"ottingen, Germany
              \email{pkaepyl@uni-goettingen.de}
   \and ReSoLVE Centre of Excellence, Department of Computer Science, Aalto University, PO Box 15400, FI-00076 Aalto, Finland
   \and Nordita, KTH Royal Institute of Technology and Stockholm University, Roslagstullsbacken 23, SE-10691 Stockholm, Sweden
   \and Department of Astronomy, Stockholm University, SE-10691 Stockholm, Sweden
   \and JILA and Department of Astrophysical and Planetary Sciences, Box 440, University of Colorado, Boulder, CO 80303, USA
   \and Laboratory for Atmospheric and Space Physics, 3665 Discovery Drive, Boulder, CO 80303, USA
   \and Max-Planck-Institut f\"ur Sonnensystemforschung, Justus-von-Liebig-Weg 3, D-37077 G\"ottingen, Germany}

   \date{$ $Revision: 1.225 $ $}

   \abstract{
     Turbulent diffusion of large-scale flows and magnetic fields plays a major
     role in many astrophysical systems, such as stellar convection zones and
     accretion discs.
   }{%
     Our goal is to compute turbulent viscosity and magnetic diffusivity
     which are relevant for diffusing large-scale flows and magnetic fields,
     respectively. We also aim to compute their ratio, which is the
     turbulent magnetic Prandtl
     number, $\Pmt$, for isotropically forced homogeneous turbulence.
   }%
   {
     We used simulations of forced turbulence in fully periodic cubes
     composed of isothermal gas with an imposed large-scale sinusoidal shear flow.
     Turbulent viscosity was computed either from the
     resulting Reynolds stress or from the decay rate of the
     large-scale flow. Turbulent magnetic diffusivity was
     computed using
     the test-field method for a microphysical magnetic Prandtl
     number of unity.
     The scale dependence of the coefficients was studied by varying
     the wavenumber of the imposed sinusoidal shear and test fields.
   }%
   {
     We find that turbulent viscosity and magnetic diffusivity are in
     general of the same order of magnitude. Furthermore, the
     turbulent viscosity depends on the fluid
     Reynolds number ($\Rey$) and scale separation ratio of turbulence. The
     scale dependence of the turbulent viscosity is found to be well approximated
     by a Lorentzian. These results are similar to those
     obtained earlier for the turbulent magnetic diffusivity. The
     results for the turbulent transport coefficients appear to
     converge at
     sufficiently high values of $\Rey$ and the scale
     separation ratio. However, a weak trend is found
     even at the largest values of $\Rey$, suggesting that the
     turbulence is not in the fully developed regime. The turbulent
     magnetic Prandtl number converges to a value that is slightly below
     unity for large $\Rey$. For small $\Rey$ we find values between 0.5 and
     0.6 but the data are insufficient to
     draw conclusions regarding asymptotics.
     We demonstrate that our results are independent of the correlation
     time of the forcing function.
   }%
   {
     The turbulent magnetic diffusivity is, in general, consistently higher than
     the turbulent viscosity, which is in qualitative
     agreement with analytic theories. However, the actual value of $\Pmt$ found
     from the simulations ($\approx0.9\ldots0.95$) at large $\Rey$
     and 
     large
     scale separation ratio is higher than any of the
     analytic predictions
      (0.4 \ldots 0.8).
   }%


   \maketitle


\section{Introduction}

Turbulent transport is often invoked to explain phenomena in
astrophysical systems such as accretion \citep[e.g.][]{SS73,FKR02},
maintenance of stellar differential rotation \citep{R80,R89,RKH13}, and
large-scale magnetic field generation \citep{M78,KR80}. Turbulence is
typically thought to diffuse large-scale structures analogously to
molecular diffusion but at a rate that is several orders of magnitude
higher \citep[e.g.][]{VBMKM14}.

Turbulent diffusion coefficients, such as turbulent viscosity ($\nut$)
and magnetic diffusivity ($\etat$), are often estimated using arguments from the mixing
length theory (hereafter MLT) according to which
$\nut\approx\etat\approx u l/3$, where $u$ and
$l$ are the characteristic velocity and the length scale of
turbulence. Such estimates yield values of the order of
$10^8\ldots10^9\m^2\s^{-1}$ for the solar convection zone, which
coincide with values estimated for the turbulent magnetic diffusivity $\etat$
from sunspot decay in the quenched case \citep{KR75,PvD-G97,RK00} and from
cross helicity measurements in the unquenched (quiet Sun) case \citep{RKB11}.
With the advent of the test-field method
\citep{SRSRC05,SRSRC07}, it has become possible to measure turbulent
transport coefficients that are relevant for the electromotive force
(e.g., the turbulent magnetic diffusivity) from simulations. Detailed
studies using
this
 method
indicate that the MLT
estimate yields the correct order of magnitude in the kinematic regime
\citep[e.g.][]{SBS08,KKB09a}, provided that $l$ is identified with
the inverse of the wavenumber $\kf$ of the energy-carrying eddies.
This result can further be affected by other physical properties, such
as the presence of kinetic helicity in the flow, which can reduce the
value of $\etat$ \citep{2017AN....338..790B}.
The test-field method also revealed an approximately Lorentzian
dependence on the wavenumber of the mean field \citep{BRS08}.

In the absence of a corresponding test-field method for hydrodynamics,
the estimates of $\nut$ are typically much less accurate than those
obtained for $\etat$ from such methods.
Estimates of turbulent viscosity from
shearing box simulations, however, also indicate a value of the order
of the MLT estimate \citep[e.g.][]{SKKL09,KBKSN10}.
Computing $\nut$ from simulations with imposed linear shear flows is
problematic due to hydrodynamical instabilities that can be excited
\citep[e.g.][]{EKR03,KMB09}. Furthermore, also non-diffusive
contributions to the
turbulent
stress exist. First, the
anisotropic kinetic alpha (AKA) effect
can occur in the presence of 
Galilean non-invariant flows
\footnote{More precisely one had to speak about flows whose statistical properties are
Galilean invariant or non-invariant.}
and can give rise to hydrodynamic instabilities analogous to the electromagnetic dynamo
\citep[e.g.][]{FSS87,BvR01,2018A&A...611A..15K}.$\!$\footnote{In those
papers, the presence of a deterministic forcing function made the flow
Galilean non-invariant.} Second, anisotropic
turbulence with global rotation leads to a $\Lambda$ effect,
which is relevant for causing differential rotation
\citep[e.g.][]{R89,KR05,KB08,Kap19}. Typically, these
effects cannot easily be
disentangled from the contribution of turbulent viscosity.
Additionally, a
spatially non-uniform kinetic helicity
\cite{2016PhRvE..93c3125Y} in rotating non-mirror symmetric flows
leads to the generation of
large-scale flows.

Contrary to the microphysical magnetic Prandtl number, which can vary
over
tens of orders of magnitude in the astrophysical context, depending on
the physical characteristics of the system under study
\citep[e.g.][]{BS05}, the ratio of $\nut$ to $\etat$,
that is
the turbulent magnetic Prandtl number $\Pmt$, is thought to be of the
order of unity in the astrophysically relevant regime of high Reynolds
numbers.
Nevertheless, astrophysical applications of the possibility of $\Pmt$
being different from unity have been discussed.
These include both accretion disc turbulence and solar convection.
In the context of accretion onto a magnetised star, one often assumes
that the field lines of the star's magnetic field are being dragged with
the flow towards the star, so as to achieve a pitch angle suitable for
jet launching \citep{1982MNRAS.199..883B}.
This requires the turbulent magnetic diffusivity to be small
\citep{2000A&A...358..612E}, 
while
 subsequent work has shown that
$\Pmt$ only has a weak influence on the pitch angle
\citep{2002A&A...393L..81R}.

Another application has been suggested in the context of the solar
convection zone.
For flux transport dynamos to explain the equatorward migration of the
sunspot belts, one must assume the turbulent magnetic diffusivity to be
of the order of $10^7\m^2\s^{-1}$ \citep{CNC04}.
On the other hand, to prevent the contours of constant angular velocity
from being
constant on cylinders, the turbulent viscosity must be around
$10^9\m^2\s^{-1}$, or even larger \citep{BTMR90}.
Thus, again, a turbulent magnetic Prandtl number in excess of unity is
required for this model to be successful.
A large turbulent viscosity is sometimes argued to be a consequence of the
magnetic stress from small-scale dynamo action \citep{2018arXiv180100560K}.
Whether this idea has a solid foundation remains open, however.

The analytic values of the turbulent magnetic Prandtl number
range between 0.4 under the first-order
smoothing approximation (hereafter FOSA) to 0.8 under various versions
of the $\tau$ approximation \citep{YBR03,KPR94}, of which the spectral
minimal $\tau$ approximation (hereafter MTA) applied to fully
developed turbulent convection yields values in the range 0.23--0.46
\citep{2006GApFD.100..243R}.
Different renormalisation group
analyses yield $\Pmt\approx0.42\ldots0.79$
\citep[e.g.][]{FSP82,KR94,2001PhPl....8.3945V,2011PhRvE..84d6311J}. Furthermore,
the turbulent magnetic
Prandtl number has been studied from simulations of forced
turbulence with a decaying large-scale field component
by \cite{YBR03} who found that $\Pmt$ is approximately
unity irrespective of the microphysical magnetic Prandtl and Reynolds
numbers. However, their dataset is limited to a few representative
cases that do not probe the Reynolds number or scale dependences systematically.

Our aim is to compute the turbulent viscosity and turbulent magnetic
Prandtl number from direct simulations of homogeneous isotropically
forced turbulence where we systematically vary the Reynolds number and
scale separation ratio and compare the obtained results with analytic
results.
To achieve this, we impose a large-scale shear flow
with a harmonic profile
on the (non-rotating) flow and determine the turbulent viscosity either from
the generated Reynolds stresses or from the decay rate of the
large-scale flow.
For obtaining the turbulent magnetic diffusivity we
employ the test-field method.

\section{Model} \label{sect:model}

\subsection{Basic equations}

We model a compressible gas in a triply periodic cube with 
edge length $L$.
It obeys an isothermal equation of state defined by
$p=\css \rho$,
with pressure $p$, density $\rho$ and constant speed of sound
 $\cs$.
Hence,
we solve the continuity and Navier--Stokes equations
with both an imposed random and large-scale shear forcing
\begin{eqnarray}
\frac{D \ln \rho}{Dt} &\!=\!& - \bm\nabla \cdot {\bm U}, \\
\frac{D {\bm U}}{Dt} &\!=\!&\!-\css \bm\nabla \ln \rho \!+\! \frac{1}{\rho} \bm\nabla\!\cdot\! (2 \nu \rho \bm{\mathsf{S}}) \!+\! {\bm f} \!-\! \frac{1}{\tau}\!\left(\mean{U}_y\!-\!\mean{U}_y^{(0)}\!\right)\hat{\bm{e}}_y, \label{equ:NS}
\end{eqnarray}
$D/Dt = \pd/\pd t + {\bm U} \cdot
\bm\nabla$ is the advective time derivative, 
${\bm U}$ is the velocity,
$\nu$ is the constant kinematic viscosity, 
and $\mathsf{S}_{ij} = \onehalf (U_{i,j}+U_{j,i}) 
-
\onethird \delta_{ij} \bm\nabla\cdot{\bm U}$ is the traceless rate of
strain tensor, where the commas denote spatial derivatives.
The forcing function $\fff$ is given by
\begin{eqnarray}
\fff=f_0 N\fff_{\kkk}\mbox{Re}\left\{\exp\!\left[i\big(\kkk(t)\cdot\xxx+\phi(t)\big)\right]\right\},  
\label{equ:forcing}
\end{eqnarray}
where $\kkk(t)$ is a random wavevector and
\begin{eqnarray}
\fff_{\kkk}(t)=\big(\kkk\times\eee(t)\big)/\sqrt{\kkk^2-\big(\kkk\cdot\eee(t)\big)^2}
\end{eqnarray}
is used to produce a nonhelical transversal sinusoidal $\fff$, where $\eee(t)$
  is an
arbitrary 
random
unit vector,
not aligned with $\kkk$,
 and
$\phi(t)$ is a random phase.
$N(t)=c_{\rm s}^{3/2}(k/\delta t)^{1/2}$ is a normalisation factor,
$k=|\kkk|$, $\delta t$ is the length of the 
integration
time step
and $f_0$ is a
constant
 dimensionless scaling factor.
The quantities $\kkk$, $\eee$, and $\phi$ change at every time
step, so that the external force is 
delta-correlated (white) in time.
Numerically, we integrate the forcing
term by using the Euler--Maruyama scheme \citep{Higham01}. 
We consider models where 
$\kkk$ is within
a narrow 
shell
of 
wavevectors
with $k$ 
close to
 a chosen $\kf$,
 and determined 
such that the forcing always obeys the periodic boundary conditions. 

The last term in Eq.~\eqref{equ:NS} maintains a large-scale shear flow 
on top of the forced
background turbulence
via relaxing
the horizontally 
($xy$)
averaged part of the $y$ velocity, indicated by the
overbar, towards the 
temporally
constant profile $\mean{U}^{(0)}_y$;
$\hat{\bm e}_y$ is the unit vector in the $y$-direction.
The relaxation time scale $\tau$ is chosen to match the 
turnover time $(\urms \kf)^{-1}$ of the turbulence, where $\urms$ is
the rms value of the fluctuating velocity,
$\urms = {\left\langle(\bm U - \mean{\bm U})^2 \right\rangle}_{\!t}^{1/2}$,
with the average taken
over the full volume as
indicated by the angle brackets, 
and over the statistically steady part of the simulations, indicated by the
subscript $t$.
Our results are not sensitive to the relaxation time $\tau$ in the
range $0.1 < \tau \urms \kf < 10$ so the (arbitrary) choice $\tau \urms \kf = 1$
is justified.
We choose a simple harmonic form for
the shear flow
according to
\begin{eqnarray}
\mean{U}^{(0)}_y = U_0 \cos(k_U z),
\label{equ:impf}
\end{eqnarray}
where $U_0$ is the flow amplitude, and
$k_U = k_1, 2 k_1, \ldots, k_U^{\rm max}$, $k_1=2\pi/L$.

\subsection{Input and output quantities}

We measure density in terms of its initially uniform value $\rho_0$,
velocity in units of the sound speed $c_{\rm s}$, and length in terms
of $k^{-1}_1$. Furthermore, in the cases with the test-field method
employed,
we
choose 
a system of electromagnetic units in which 
$\mu_0=1$, where $\mu_0$ is the permeability of vacuum.
The simulations are fully defined by choosing the forcing amplitude
$f_0$ and scale $\kf/k_1$, kinematic viscosity $\nu$, microscopic
magnetic Prandtl number
\begin{eqnarray}
\Pm=\frac{\nu}{\eta},
\end{eqnarray}
where $\eta$ is the microscopic magnetic diffusivity in the test-field
method, and the shear
parameter
\begin{eqnarray}
\Sh_{\rm c}=\frac{U_0 k_U}{\cs \kf}.
\end{eqnarray}
We further assume that the scale of the test fields always equals that
of the imposed large-scale flow, 
that is
$k_B=k_U$, and that the value
of $\Pm$ for the test-field simulations equals unity.
For the scale separation ratio ${\mathscr S}$ we employ the definition
\begin{eqnarray}
{\mathscr S}=\kf/k_U.
\end{eqnarray}
The following quantities are used as diagnostics of our models.
We quantify the level of turbulence in the simulations by the fluid
and magnetic Reynolds numbers
\begin{equation}
\Rey=\frac{\urms}{\nu \kf},\ \ \ \Rm=\frac{\urms}{\eta \kf}=\Pm \, \Rey.
\end{equation}
The strength of
the imposed shear is measured by the dynamic shear number
\begin{eqnarray}
\Sh=\frac{U_0 k_U}{\urms \kf}.
\end{eqnarray}
Guided by MLT and FOSA, respectively, we normalise both
the turbulent viscosity and magnetic diffusivity by
\begin{eqnarray}
\nutz=\etatz=\urms/3\kf,
\end{eqnarray}
while the turbulent magnetic Prandtl number is given by
\begin{eqnarray}
\Pmt=\frac{\nut}{\etat}.
\end{eqnarray}

\section{Computation of $\nut$ and $\etat$}

\subsection{Mean-field analysis}

In what follows, we rely upon Reynolds averaging
specifically defining the mean quantities as averages over $x$ and $y$. Hence, 
they
can only depend on $z$ and time. 
Averages are indicated
by overbars and fluctuations by lowercase or primed
quantities,
thus ${\bm U} = \meanv{U} + \bm u$, $\rho=\mean{\rho}+\rho'$ etc.

\subsubsection{Hydrodynamics} \label{sec:hydro}

In the incompressible case all turbulent effects can be subsumed in
the Reynolds stress tensor
$Q_{ij}=\mean{u_i u_j}$
 whose
divergence appears in the evolution equation of the mean
flow. Including compressibility and starting from
\begin{align}
  &\partial_t (\rho U_i) + \partial_j (\rho U_i U_j) = -\partial_i P + \ldots,
\intertext{where the dots stand for viscous and external forces,
one obtains after averaging}
   &\partial_t (\mean{\rho' u_i} + \mean{\rho}\, \mean{U}_i) \nonumber\\
   &+ \partial_j (\mean{\rho} \,\mean{u_i u_j} + \mean{U}_i \mean{\rho' u_j}+ \mean{U}_j \mean{\rho' u_i} + \mean{\rho' u_i u_j} +  \mean{\rho}\, \mean{U}_i \mean{U}_j ) \nonumber\\
      &= -\partial_i \mean{P} + \ldots. \label{eq:divform}
\end{align}
The contributions proportional to the Reynolds stresses,
$\mean{\rho} \,\mean{u_i u_j}$, no longer
cover all turbulent effects originating from the inertial terms.
However, in our weakly compressible setups with $\Ma\approx0.1$ the
  difference between, for example $|\mean{(\rho u_y)' u'_z}/\urms^2|$ and
  $|\mean{\rho}\mean{u'_y u'_z}\urms^2|$ is ${\cal O}(10^{-2})$.
\footnote{Neglecting density fluctuations may not be rigorously
justified, given that the variety of potentially new effects owing to
compressibility has not yet been fully explored,
but see the recent studies by \cite{2018JPlPh..84e7302R}
and \cite{2018JPlPh..84e7301Y} for the electromotive force.
However, recent hydrodynamic results for the $\Lambda$ effect
  suggest that the effect of compressibility is weak up to
  $\Ma\approx0.8$ \citep{2019AN....340..744K}.}
Thus, we will consider, as
in the incompressible case, only the Reynolds
stresses.\footnote{Further turbulence effects result from the term
  $\mean{\boldsymbol{\mathsf{S}} \cdot \nab \ln \rho}$, but are not
  considered here either because of our assumption of weak
  compressibility. We recall that $\boldsymbol{\mathsf{S}}$ is the
  traceless rate of strain tensor used in Eq.~\eqref{equ:NS}.}
  When restricting to first order in the mean quantities,
  they can be decomposed into three contributions,
\EQ
Q_{ij} = Q_{ij}^{(0)}+Q_{ij}^{(\mlnrho)} + Q_{ij}^{(\overline{U})}
\label{Qijdecomp}
\EN
where $Q_{ij}^{(0)}$ is already present in the absence of both a mean flow
$\meanv{U}$ and a gradient of $\mean{\ln \rho}$, $Q_{ij}^{(\mlnrho)}$ is
occurring due to the presence of $\nab\mean{\ln \rho}$, and $Q_{ij}^{(\mUUU)}$
occurs due to the presence of $\meanv{U}$;
(for a justification see Appendix~{\ref{app:deriv}}.)
As in our
simulations no significant $\nab\mean{\ln \rho}$ occurs, we
disregard $Q_{ij}^{(\mlnrho)}$.  Further, as the fluctuations are
isotropically forced, the only non-zero components of $Q_{ij}^{(0)}$
are $Q_{xx}^{(0)}=Q_{yy}^{(0)}=Q_{zz}^{(0)}$. Apart from small
fluctuations, they do not depend on $z$ and thus do not act onto the
mean flow. Note that due to the absence of a global rotation there is
also no contribution of the $\Lambda$ effect in $Q_{ij}$.  In what
follows we drop the superscript $(\mUUU)$ for brevity.

For sufficiently slowly varying mean flows and sufficient scale separation,
$Q_{ij}$ can be approximately represented by the truncated Taylor expansion
\begin{align}
    Q_{ij} &= A_{ijk} \mean{U}_k + N_{ijkl} \mean{U}_{k,l}\;,   \label{eq:Qtrunc} \\
    \intertext{with the symmetry requirements}
    \quad A_{ijk} &= A_{jik}, \quad N_{ijkl} = N_{jikl}.
\end{align}
Here, $A_{ijk}$ describes the AKA effect, while $N_{ijkl}$ comprises
turbulent viscosity (amongst other effects).%
\footnote{Note that relation \eqref{eq:Qtrunc} is yielding the stresses without truncation when interpreted to be a representation of the Fourier-transformed kernel of a general convolution-like relationship between
$Q_{ij}$ and $\mean{U}_j$
\citep[cf.][]{BRS08}.}
For isotropic (and hence homogeneous) fluctuations, that is in the
kinematic limit
$\mUUU\rightarrow\boldsymbol{0}$,
$A_{ijk}=0$, and $N_{ijkl}$ must have the form
\EQ
N_{ijkl} = - \nut (\delta_{ik} \delta_{jl} + \delta_{il} \delta_{jk}) - \zeta_{\rm t} \delta_{ij} \delta_{kl}, \label{eq:Nrepr}
\EN
where the constants $\nut$ and $\zeta_{\rm t}$ are the turbulent shear
and bulk viscosities, respectively.
The Reynolds stresses appear then correspondingly as
\EQ
Q_{ij} = -\nut (\mean{U}_{i,j} + \mean{U}_{j,i}) - \sigma_{\rm t} \delta_{ij} \nab\cdot \meanv{U},\label{eq:Qrepr}
\EN
with the first term reproducing the Boussinesq ansatz.
Although our turbulence is isotropically forced, the presence of
finite shear causes it to be anisotropic
with preferred directions given by the direction of the mean flow $\meanv{U}$ and, say, its curl, $\meanv{W}=\nab\times\meanv{U}$.
Given that it is the divergence of $Q_{ij}$ which enters the mean momentum equation and
 mean quantities depend only on $z$, merely
 the components $A_{i3k}$ and $N_{i3k3}$ matter
  in \eqref{eq:Qtrunc}.
As $\meanv{U}$ needs not to be solenoidal, $\mean{U}_z$ might in general depend on $z$ and the turbulent bulk viscosity is then of interest.

Further simplification is obtained when assuming that the mean velocity has only one component. In our setup, the mean flow is always very close to the maintained
one, that is, $\meanv{U} \approx \meanv{U}^{(0)} \sim \eee_y$.
Then we have
\begin{align}
      Q_{i3} = A_{i32} \mean{U}_y + N_{i323} \,
      \mean{U}_{y,z} 
     = a_i \mean{U}_y + n_i \, \mean{U}_{y,z},
\end{align}
where we have introduced new coefficients $a_i= A_{i32}$ and $n_i= N_{i323}$. Comparison with \eqref{eq:Qrepr} reveals that  for $\mean{U}_y \rightarrow 0$ with isotropic forcing $n_2 \rightarrow -\nut$
while $a_i$ and $n_{1,3}$ should approach zero.

We note that the
AKA-effect can only be expected to appear in 
Galilean non-invariant
flows
\citep{FSS87}. This is not the case for
the flows considered here, because the forcing is $\delta$ correlated
in time, so there is no difference between a forcing 
defined in an arbitrary inertial frame
and a forcing defined in the (resting) lab frame.

\subsubsection{Magnetohydrodynamics}
\label{sec:MHD}
We consider only $z$ dependent mean fields in which case the mean
electromotive force $\mEMF$,
when truncated in analogy to \eqref{eq:Qtrunc}, can be represented
by two rank--2 tensors
\begin{align}
   \EMFi= \alpha_{ij} \mean{B}_j - \eta_{ij} \mean{J}_j, \quad \meanv{J}  =  \nab\times\meanv{B}, \quad i,j=1,2,3,
   \label{eq:ranktwoemf}
\end{align}
where $\JJJ=\bm\nabla \times\BBB$ is the current density.
Given that all quantities depend only on $z$, we have $\mean{J}_z=0$ and
because $\mean{B}_z =$ const. by virtue of $\nab\cdot\meanv{B}=0$,
 $\mEE_z(z)$
has no effect on the evolution of $\meanv{B}$. Hence we set $\mean{B}_z =0$ and restrict our interest to the components $\alpha_{ij}$ and $\eta_{ij}$ with $i,j =1,2$.
As the pseudo-tensor $\alpha_{ij}$ can for non-helical forcing merely
be constructed from the building blocks $\mean{U}_i$ and $\mean{W}_j$
by the products $\mean{U}_i \mean{W}_j$ and $\mean{U}_j \mean{W}_i$,
within its restricted part,
only the components $\alpha_{12}$ and $\alpha_{21}$ can be non-zero for our setup.
Building blocks for the anisotropic part of the restricted $\eta_{ij}$ are here
\begin{align}
   \mean{W}_i \mean{W}_j,  \mean{U}_i \mean{U}_j, \quad \text {and higher order terms},
\end{align}
hence the off-diagonal components $\eta_{12,21}$ need to vanish.
So all the relevant components,
except an isotropic contribution to $\boldsymbol{\eta}$,
have leading order in $U_0$ of at least 2.
In the limit $U_0\rightarrow 0$ we have $\alpha_{ij} \rightarrow 0$
while
$\eta_{11,22} \rightarrow \etat$.

\subsection{Imposed shear method}
\label{sec:impshear}

We apply three methods to extract the mean-field coefficients from the
simulation data \\[1mm]
{\bf M1:} \parbox[t]{.92\linewidth}{The mean flow $\mean{U}_y$ depends on $z$, and as it is
  approximately harmonic, its zeros do not coincide with those of
  its derivative $\mean{U}_{y,z}=-\mean{W}_x$.
  Hence the coefficients $a_i$ and $n_i$ can
  be isolated by
  \begin{alignat}{3}
     &a_i\!\left(\zVj\!,t\right) &&=&   &Q_{iz}\!\left(\zVj\!,t\right)\!/\mean{U}_y\!\left(\zVj\!,t\right), \label{equ:ai} \\
     &n_i\!\left(\zUj\!,t\right) &&=& \, -&Q_{iz}\!\left(\zUj\!,t\right)\!/\mean{W}_x\!\left(\zUj\!,t\right), \label{equ:ni}
  \end{alignat}
  where $\zUj$ and $\zVj$ are the zeros of $\mean{U}_y$ and
  $\mean{W}_x$, respectively.
 $a_i$ and $n_i$ are then further subjected to temporal averaging.
In general, their values at the different zeros will only
  coincide in the limit $U_0 \rightarrow 0$, but in our case the
  differences turned out to be smaller than the error bars.}
  \\[3mm]
 {\bf M2a:} \parbox[t]{.907\linewidth}{We use constant fit coefficients $a_i$ and $n_i$ in the time
  averaged simulation data of $Q_{ij}$, $\mean{U}_y$, and $\pd_z
  \mean{U}_y$:
\begin{eqnarray}
    Q_{iz} = a_i \mean{U}_y + n_i \pd_z \mean{U}_y.
    \label{equ:qijfit}
  \end{eqnarray}}
  {\bf M2b:} \parbox[t]{.907\linewidth}{Alternatively, we drop the non-diffusive contribution and use
  only a single coefficient $n_i$ as a fit parameter:
  \begin{eqnarray}
    Q_{iz} = n_i \pd_z \mean{U}_y.
    \label{equ:bouan}
  \end{eqnarray}}

\noindent
For method M1 we divide the time series of $a_i$ and $n_i$ into three
parts and
 define
 the
largest deviation from the average, taken over the whole time series. as
the error. For M2a,b we similarly perform the fit for data averaged
over three equally long parts of the time series and take the error to be
the largest deviation from the fitted values obtained from a time
average over the full time series.
Our results indicate that only the Reynolds stress component
$\qyz$ shows a significant signal that can be related to the
mean-field effects discussed above.

\begin{figure}[t]
\centering
\includegraphics[width=\columnwidth]{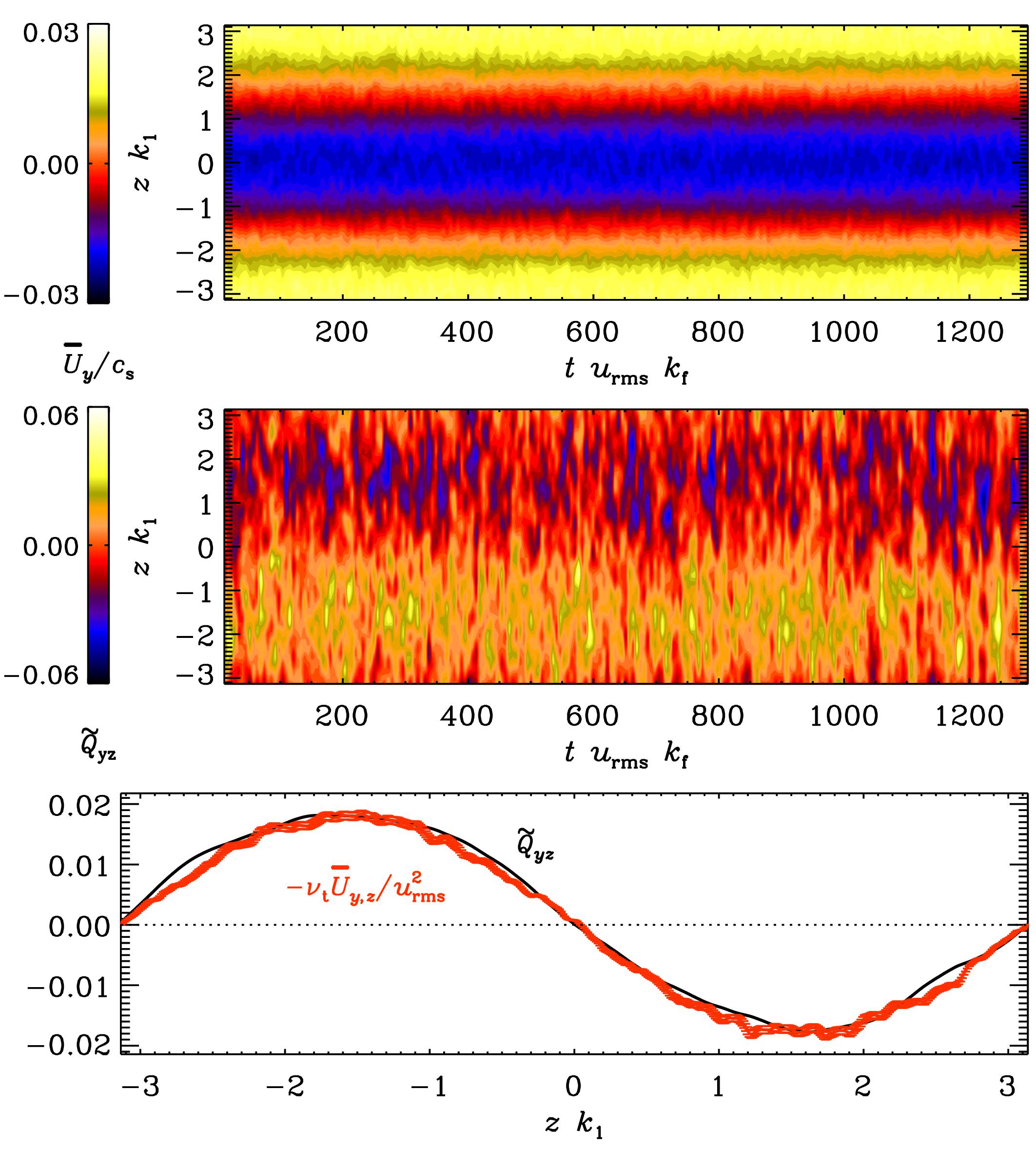}
\caption{Horizontally averaged velocity $\mean{U}_y(z,t)$ 
(top),
  $\tilde{Q}_{yz}
  (z,t)
  =\qyz/\urms^2$ (middle), and
  its temporal average
  in comparison with $-\nut \mean{U}_{y,z}/\urms^2$
  (bottom) from Run~E9
  (see Table~\ref{tab:runs2})
  with $\tilde{k}_{\rm
    f}=\kf/k_1=5$, $\tkU=k_U/k_1=1$, $\Sh\approx0.04$, and
  $\Rey\approx497$.
  $\nut$ from method M2b.
  }
\label{fig:pprofs}
\end{figure}

Figure~\ref{fig:pprofs} shows the horizontally averaged mean flow
$\mean{U}_y(z,t)$, Reynolds stress component $\qyz(z,t)$, and the
$z$ profiles
of
its temporal average along with
$-\nut \mean{U}_{y,z}$ from method M2b. The
imposed velocity profile induces a large-scale pattern in
the Reynolds stress with the same vertical wavenumber, but with a
vertical shift of $\pi/2$.

\subsection{Decay experiments} \label{sec:decay}

Apart from measuring the response of the system to imposed shear, it
is possible to measure the turbulent viscosity independently from the
decay of large-scale flows. We refer to this procedure as M3. We
employ this method to check the consistency of methods M1 and M2 in a
few cases.

The dispersion relation for the large-scale flow
$\mean{U}_y$ is
given by
\begin{equation}
\omega=-\nu_{\rm T} k_z^2,
\label{equ:disp}
\end{equation}
where $\nu_{\rm T}=\nu+\nut$ and $k_z$ is the wavenumber of the flow.
Equation~(\ref{equ:disp}) is valid if large-scale velocities other than
$\mean{U}_y$, the pressure gradient, and the effects of
compressibility are negligible. We measure the decay rate of the
$k_z=k_U$
constituent
of the flow by extracting its amplitude using
Fourier transform and fitting an exponential function to the data.
The clear exponential decay is drowned by the random signal from the
turbulence after a time that depends on the amplitude of the initial
large-scale flow and other characteristics of the simulations. Thus we
limit the fitting to the clearly decaying part of the time series
which typically covers roughly 300 turnover times.

\begin{figure}[t]
\centering
\includegraphics[width=\columnwidth]{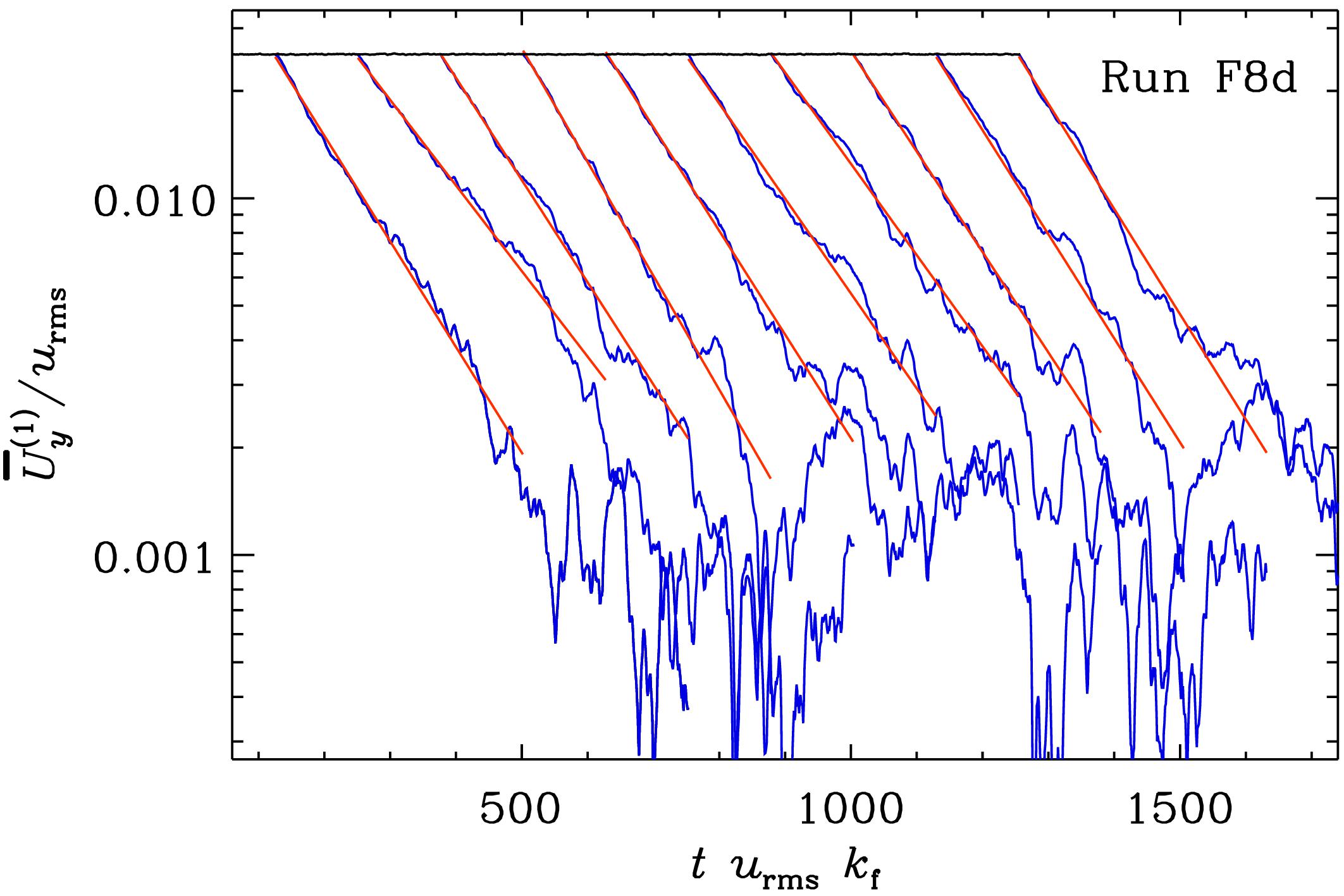}
\caption{Amplitude of the $k=k_1$ constituent
of $\mean{U}_y(z,t)$ normalised by $\urms$ in
  Run~F8d as a function of time from ten independent realisations of
  the
  decay. The solid red lines show exponential fits to the data.}
\label{fig:plot_decay}
\end{figure}

To reduce
 the effect of the stochastic fluctuations of the turbulence,
 we perform $N$ independent
realisations of the decay and measure $\nut$ from the decay rate in
each case. This is achieved by using $N$ uncorrelated snapshots from
the fiducial run
with imposed shear flow as initial conditions for
decay
experiments, see \Fig{fig:plot_decay} for representative
results where $N=10$. Such snapshots are separated by at least 80 turbulent eddy
turnover
times. An error estimate is obtained by dividing the obtained values
of $\nut$ into two groups and considering the largest deviation of
averages over these from the average over the full set.

\begin{table}[t!]
\centering
\caption[]{Summary of the runs with varying shear.}
      \label{tab:runs1}
      \vspace{-0.5cm}
     $$
         \begin{array}{p{0.05\linewidth}ccrrccccccc}
           \hline
           \noalign{\smallskip}
Run & \Rey  & \Sh & \Sh_{\rm c} [10^{-5}]\!\!\!\! & \tkf \!  & \Ma & \tnut & \tetat & \Pmt & \mbox{Grid} \\ \hline
A1 &  21 &  0.015 &   160 &  3  &  0.105 & 1.509 &  1.899 & 0.795 &  72^3 \\ 
A2 &  21 &  0.030 &   319 &  3  &  0.106 & 1.566 &  1.873 & 0.836 &  72^3 \\ 
A3 &  21 &  0.060 &   638 &  3  &  0.106 & 1.575 &  1.912 & 0.824 &  72^3 \\ 
A4 &  22 &  0.146 &  1595 &  3  &  0.109 & 1.452 &  1.900 & 0.764 &  72^3 \\ 
A5 &  25 &  0.264 &  3191 &  3  &  0.121 & 1.496 &  2.098 & 0.713 &  72^3 \\ 
A6 &  35 &  0.369 &  6381 &  3  &  0.173 & 1.618 &  2.192 & 0.738 &  72^3 \\ 
\hline
B1 &  21 &  0.009 &    98 &  5  &  0.105 & 1.768 &  2.018 & 0.876 & 144^3 \\ 
B2 &  21 &  0.019 &   196 &  5  &  0.105 & 1.699 &  2.031 & 0.836 & 144^3 \\ 
B3 &  21 &  0.037 &   392 &  5  &  0.106 & 1.769 &  2.097 & 0.844 & 144^3 \\ 
B4 &  21 &  0.091 &   981 &  5  &  0.107 & 1.760 &  2.083 & 0.845 & 144^3 \\ 
B5 &  23 &  0.171 &  1962 &  5  &  0.115 & 1.818 &  2.361 & 0.770 & 144^3 \\ 
B6 &  33 &  0.235 &  3923 &  5  &  0.167 & 2.180 &  4.048 & 0.538 & 144^3 \\ 
\hline
C1 &  21 &  0.005 &    50 &  10 &  0.106 & 1.856 &  2.110 & 0.880 & 144^3 \\ 
C2 &  21 &  0.009 &   100 &  10 &  0.106 & 1.905 &  2.113 & 0.901 & 144^3 \\ 
C3 &  21 &  0.019 &   199 &  10 &  0.106 & 1.916 &  2.136 & 0.897 & 144^3 \\ 
C4 &  21 &  0.047 &   499 &  10 &  0.106 & 1.926 &  2.216 & 0.869 & 144^3 \\ 
C5 &  22 &  0.092 &   997 &  10 &  0.109 & 1.969 &  2.450 & 0.804 & 144^3 \\ 
C6 &  23 &  0.131 &  1496 &  10 &  0.115 & 2.063 &  2.811 & 0.734 & 144^3 \\ 
C7 &  29 &  0.136 &  1994 &  10 &  0.147 & 3.140 &  5.037 & 0.623 & 144^3 \\ 
\hline
         \end{array}
     $$
         \tablefoot{
         $\Ma=\urms/c_{\rm s}$ is the Mach number,
           $\tnut=\nut/\nutz$, and $\tetat=\etat/\etatz$. Furthermore,
            $\tkB=\tkU=1$
           in all
           runs yielding ${\mathscr S}=\tkf$.
           Here and in subsequent tables, $\nut$ is measured 
           with
             method M2b and $\etat$ 
            with
            the test-field method.}
\end{table}

\begin{figure*}[t]
\centering
\includegraphics[width=\textwidth]{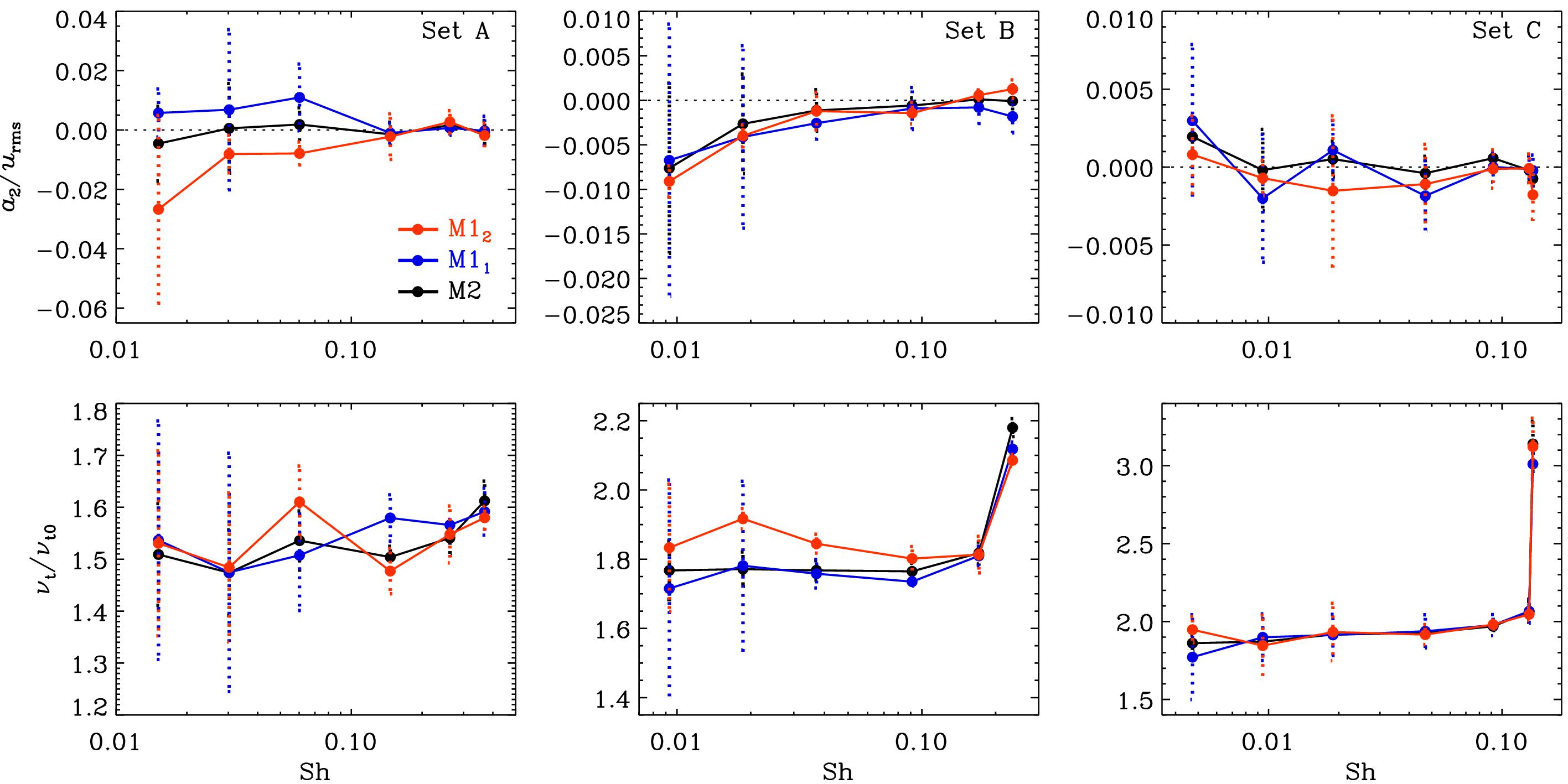}
\caption{AKA--effect coefficient $a_2$ ({\it top })
and turbulent
  viscosity $\nut$ ({\it bottom})
  as functions of $\Sh$ for three
  scale separation ratios, 
  $\mathscr S=3$
   (Set~A, left), 5 (Set~B, middle),
  and 10 (Set~C, right). The colours refer to methods M1 (blue and
  red), and M2 (black). M1$_1$ and M1$_2$ refer to the two zeros from
  \Eqsa{equ:ai}{equ:ni}.}
\label{fig:pakanut}
\end{figure*}

\begin{figure}[t]
\centering
\includegraphics[width=\columnwidth]{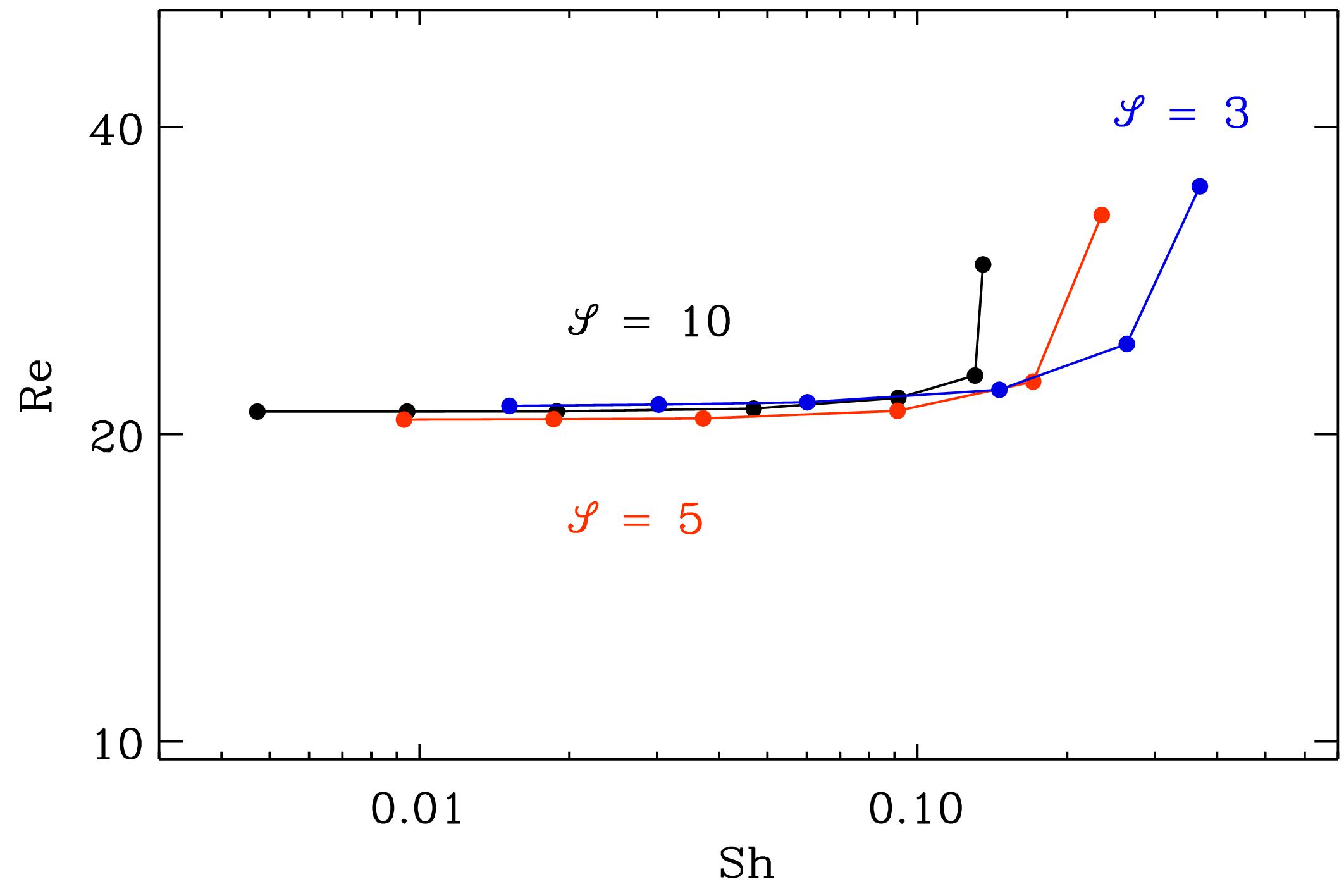}
\caption{Reynolds number as a function of $\Sh$ for three scale
  separation ratios, 
  $\mathscr S=3$
  (blue), 5 (red), and 10 (black), or
  Sets~A, B, and C,
  respectively.}
\label{fig:pSh}
\end{figure}

\begin{figure*}[t]
\centering
\includegraphics[width=0.96\textwidth]{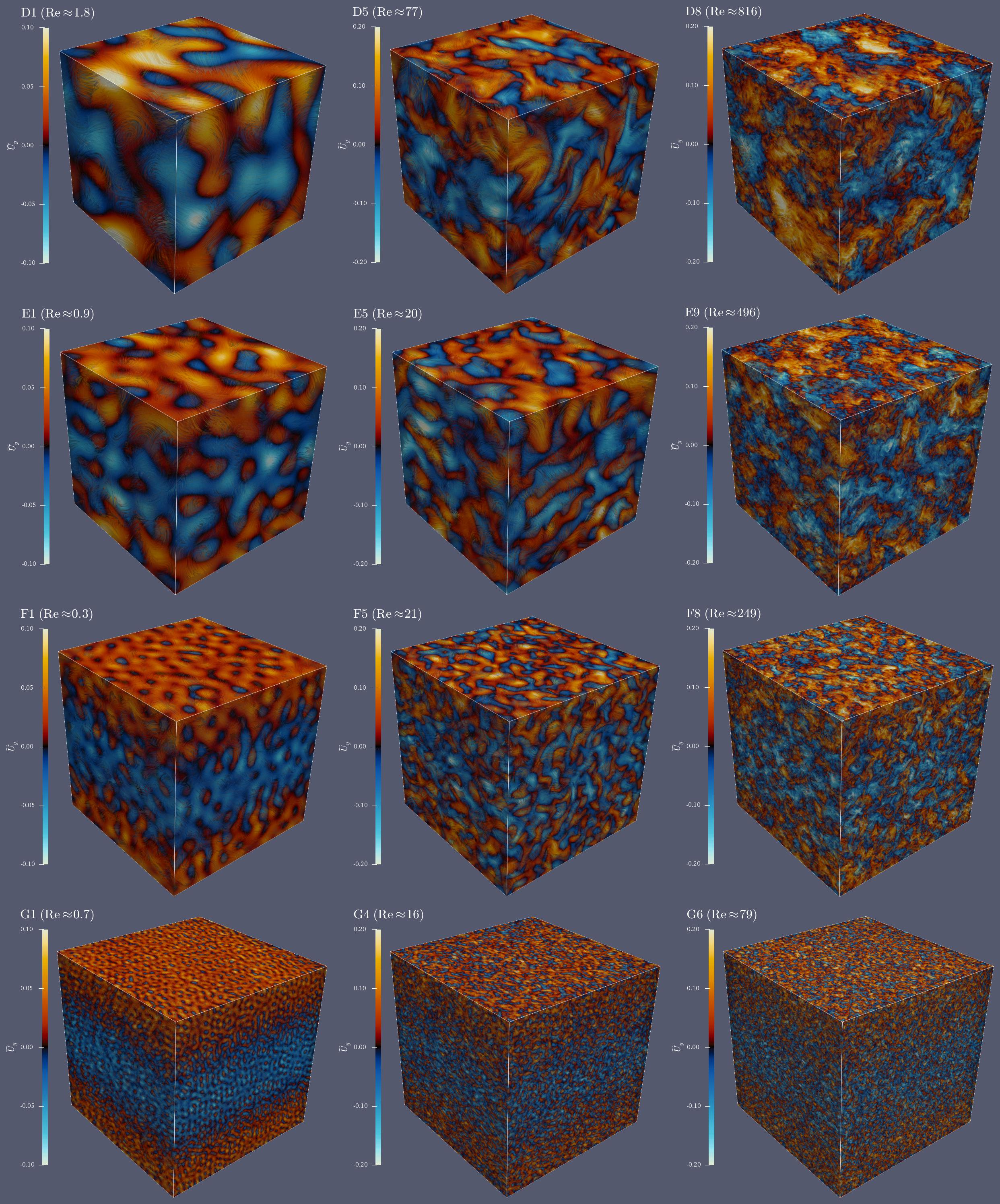}
\caption{
Normalised
streamwise velocity component $\tilde{U}_y=U_y/\cs$ at the periphery of the
  computational domain for increasing scale separation ratio (from top
  to bottom), and increasing Reynolds number (left to right)
  for selected runs from D1 to G6.}
\label{fig:boxes}
\end{figure*}

\subsection{Test-field method}

We use the test-field method, originally described in
\cite{SRSRC05,SRSRC07}, to determine the turbulent transport
coefficients
$\alpha_{ij}$ and $\eta_{ij}$.
Our formulation is
essentially the same as in \cite{BRRK08}. The
fluctuating magnetic fields
\begin{eqnarray}
\frac{\pd \aaaT}{\pd t} = \mUUU \times \bbbT + \uuu \times \mBBBT + \left(\uuu \times \bbbT\right)' + \eta \bm\nabla^2 \aaaT,
\end{eqnarray}
are evolved with the flow taken from the simulation (main run), where
$\bbbT=\bm\nabla\times\aaaT$,
$\eta$ is the magnetic diffusivity,
and $\meanv{B}^{\rm T}$ is one out of a set of
large-scale test fields.
Neither the
fluctuating fields $\aaaT$ nor the test fields $\BBBT$ act back
on the flow. Each of the test fields yields an electromotive force
(EMF)
\begin{eqnarray}
\mEMF = \overline{\uuu \times \bbbT}.
\end{eqnarray}
Assuming that the mean field $\mBBB$ varies slowly in space and time,
the electromotive force can be written as
\begin{eqnarray}
\EMFi = \alij \mBBj + \beijk \frac{\pd \mBBk}{\pd x_j},\label{eq:EMFi}
\end{eqnarray}
where $\alij$ and $\beijk$ represent the $\alpha$ effect and turbulent
diffusion, respectively. These coefficients can be
unambiguously
 inverted from
Eq.~(\ref{eq:EMFi}) by choosing an appropriate number of independent
test fields.

We use
four
stationary $z$ dependent test fields
\begin{equation}
\begin{alignedat}{6}
&\mBBB^{1c}\!&&= && \,B_0(\cos k_B z,0,0), \ \
&&\mBBB^{2c}\!&&= && \,B_0(0,\cos k_B z,0), \\
&\mBBB^{1s}\!&&= && \,B_0(\sin k_B z,0,0), \ \
&&\mBBB^{2s}\!&&= && \,B_0(0,\sin k_B z,0),
\end{alignedat}
\end{equation}
where $k_B$ is a wavenumber.
As explained in \Sec{sec:MHD}, \Eq{eq:EMFi} simplifies here to \Eq{eq:ranktwoemf} with
$\eta_{i1}=\beta_{i23}$ and $\eta_{i2}=-\beta_{i13}$.
Because of $\alpha_{11,22}=0, \;\alpha_{12,21}\rightarrow 0$ for $U_0 \rightarrow 0$
we neglect the latter for weak imposed shear flows 
 and 
 simplify
 \Eq{eq:ranktwoemf} 
 further to
\begin{eqnarray}
\EMFi = -\etij \mJJj.\label{eq:EMFi3}
\end{eqnarray}
We are interested in the diagonal components of $\etij$ which we
represent in terms of turbulent diffusivity by
\begin{eqnarray}
\etat = \onehalf (\eta_{11} + \eta_{22}).
\end{eqnarray}
In the case of homogeneous isotropic turbulence, the turbulent
transport coefficients are uniform across the system
and
volume averages are
appropriate. In the present case,
however,
 the turbulence 
 can neither
 be considered
fully isotropic 
nor homogeneous
due to the imposed 
$z$ dependent
shear flow,
which makes the coefficients also anisotropic and $z$ dependent.
Both effects are weak though
in the computed $\etat$;
 see
  \Seca{sec:tudi}
for the effect of anisotropy.

Exponential growth of the test solutions $\bbbT$ at high
$\Rm$ is a known
issue in the test-field method \citep{SBS08}.
To circumvent it,
we reset the fluctuating
fields $\bbbT$ periodically
to zero
with a resetting time that is roughly
inversely proportional to the magnetic Reynolds number.
The error of the turbulent magnetic Prandtl number is computed from
\begin{eqnarray}
\delta \Pmt \approx \Pmt \left(\frac{\delta\nut}{\nut} + \frac{\delta\etat}{\etat}\right),
\end{eqnarray}
where $\delta\nut$ and $\delta\etat$ are the errors of the turbulent
viscosity and diffusivity, respectively.

\begin{figure}[t]
\centering
\includegraphics[width=\columnwidth]{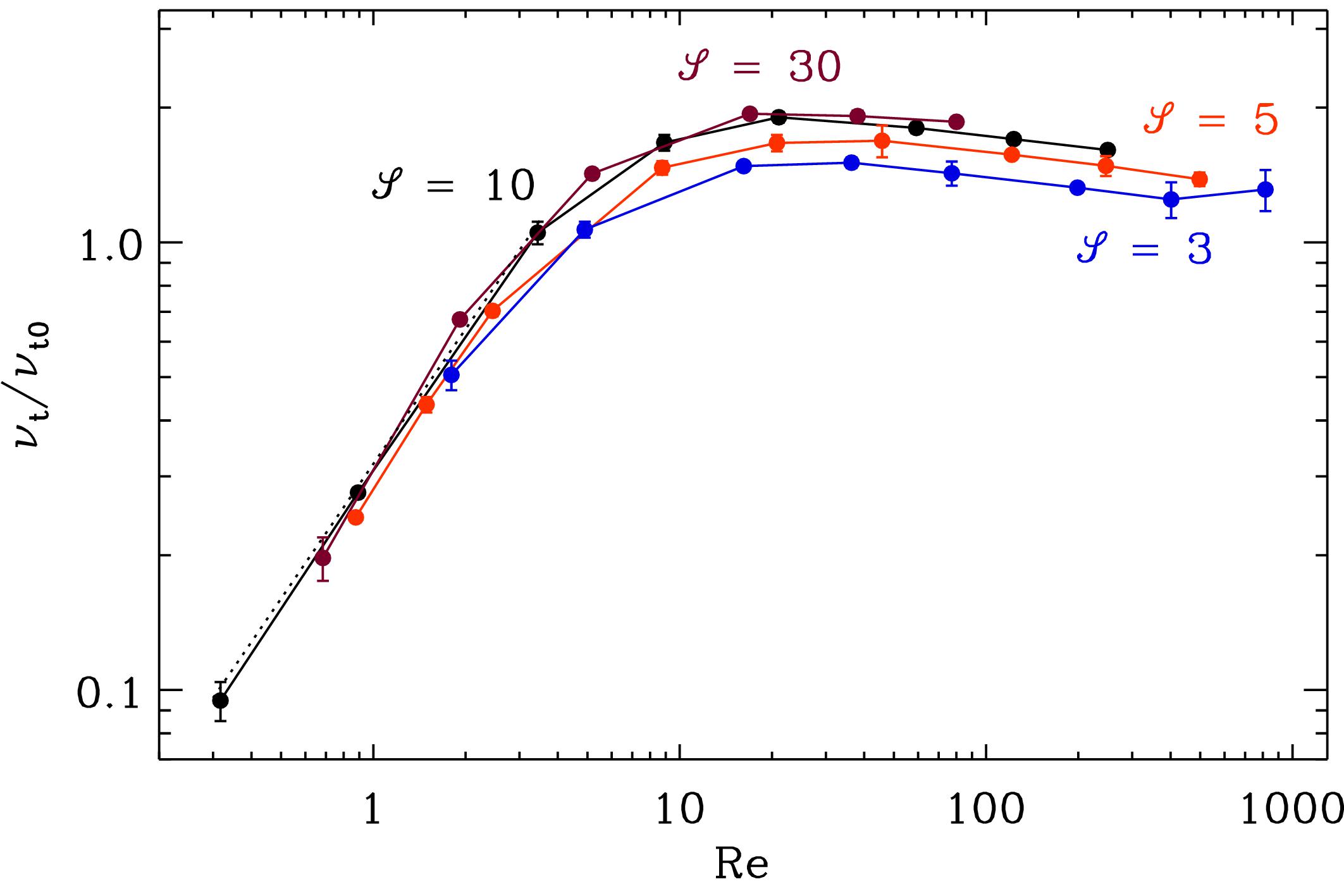}
\caption{Turbulent viscosity $\nut$,
 normalised by $\nutz$,
  as a function of the Reynolds
  number $\Rey$ 
  for sets of runs with different scale separation ratio $\mathscr{S}$,
  but
  $\Sh_{\rm c}=$const.\
  within each of the Sets
  D (blue),
  E (red), F
  (black), and G (purple). The dotted black line is proportional to
  $\Rey$.}
\label{fig:pnut_Re}
\end{figure}

\begin{table}[t!]
\centering
\caption[]{Summary of the runs with varying Reynolds numbers.}
      \label{tab:runs2}
      \vspace{-0.5cm}
     $$
         \begin{array}{p{0.05\linewidth}cccccccccccc}
           \hline
           \noalign{\smallskip}
           Run & \Rey & \Sh & \Sh_{\rm c} [10^{-5}] & \tkf  & \Ma & \tnut & \tetat & \Pmt & \mbox{Grid} \\ \hline
           D1 & 1.8 & 0.144 &  813 &  3  & 0.056 & 0.505 & 0.857 & 0.589 &  64^3 \\ 
           D2 & 4.9 & 0.106 &  813 &  3  & 0.077 & 1.068 & 1.594 & 0.670 &  64^3 \\ 
           D3 &  16 & 0.080 &  813 &  3  & 0.101 & 1.481 & 1.896 & 0.781 &  64^3 \\ 
           D4 &  36 & 0.071 &  813 &  3  & 0.114 & 1.508 & 1.832 & 0.823 &  64^3 \\ 
           D5 &  77 & 0.067 &  813 &  3  & 0.121 & 1.430 & 1.735 & 0.824 & 128^3 \\ 
           D6 & 198 & 0.065 &  813 &  3  & 0.124 & 1.329 & 1.590 & 0.836 & 128^3 \\ 
           D7 & 401 & 0.065 &  813 &  3  & 0.126 & 1.247 & 1.551 & 0.804 & 256^3 \\ 
           D8 & 816 & 0.064 &  813 &  3  & 0.128 & 1.310 & 1.489 & 0.880 & 512^3 \\ 
           \hline
           E1 & 0.9 & 0.112 &  500 &  5  & 0.045 & 0.242 & 0.447 & 0.540 &  64^3 \\ 
           E2 & 1.5 & 0.094 &  500 &  5  & 0.053 & 0.434 & 0.762 & 0.570 &  64^3 \\ 
           E3 & 2.4 & 0.080 &  500 &  5  & 0.062 & 0.702 & 1.182 & 0.594 &  64^3 \\ 
           E4 & 8.8 & 0.056 &  500 &  5  & 0.089 & 1.468 & 2.007 & 0.732 &  64^3 \\ 
           E5 &  20 & 0.047 &  500 &  5  & 0.106 & 1.665 & 2.020 & 0.824 &  64^3 \\ 
           E6 &  45 & 0.043 &  500 &  5  & 0.117 & 1.693 & 1.901 & 0.891 & 128^3 \\ 
           E7 & 121 & 0.040 &  500 &  5  & 0.124 & 1.578 & 1.733 & 0.911 & 256^3 \\ 
           E8 & 246 & 0.040 &  500 &  5  & 0.125 & 1.485 & 1.678 & 0.885 & 256^3 \\ 
           E9 & 497 & 0.039 &  500 &  5  & 0.127 & 1.382 & 1.604 & 0.862 & 512^3 \\ 
           \hline
           F1 & 0.3 & 0.080 &  254 &  10 & 0.032 & 0.096 & 0.156 & 0.617 & 128^3 \\ 
           F2 & 0.9 & 0.057 &  254 &  10 & 0.045 & 0.274 & 0.454 & 0.603 & 128^3 \\ 
           F3 & 3.4 & 0.037 &  254 &  10 & 0.069 & 1.049 & 1.532 & 0.685 & 128^3 \\ 
           F4 & 8.9 & 0.028 &  254 &  10 & 0.089 & 1.670 & 2.103 & 0.794 & 128^3 \\ 
           F5 &  21 & 0.024 &  254 &  10 & 0.106 & 1.905 & 2.116 & 0.900 & 128^3 \\ 
           F6 &  59 & 0.021 &  254 &  10 & 0.119 & 1.787 & 1.926 & 0.928 & 256^3 \\ 
           F7 & 123 & 0.021 &  254 &  10 & 0.123 & 1.700 & 1.802 & 0.943 & 512^3 \\ 
           F8 & 249 & 0.020 &  254 &  10 & 0.125 & 1.607 & 1.712 & 0.939 & 512^3 \\ 
           \hline
           G1 & 0.7 & 0.021 &   84 &  30 & 0.041 & 0.205 & 0.356 & 0.577 & 288^3 \\ 
           G2 & 1.9 & 0.015 &   84 &  30 & 0.058 & 0.675 & 1.114 & 0.606 & 288^3 \\ 
           G3 & 5.2 & 0.011 &   84 &  30 & 0.078 & 1.429 & 2.090 & 0.684 & 288^3 \\ 
           G4 &  16 & 0.008 &   84 &  30 & 0.102 & 1.930 & 2.215 & 0.871 & 288^3 \\ 
           G5 &  38 & 0.007 &   84 &  30 & 0.114 & 1.915 & 2.072 & 0.924 & 576^3 \\ 
           G6 &  79 & 0.007 &   84 &  30 & 0.120 & 1.856 & 1.844 & 1.007 & 576^3 \\ 
           \hline
         \end{array}
     $$
         \tablefoot{All quantities have the same meanings as in Table~\ref{tab:runs1}.
           Again $\tkB=\tkU=1$ and ${\mathscr S}=\tkf$.
         }
\end{table}

\section{Results} \label{sect:results}

We perform several sets of simulations where we vary the forcing
wavenumber $\kf$, determining the scale separation ratio, fluid and
magnetic Reynolds numbers $\Rey$ and $\Rm$, respectively, and the wavenumber of the
large-scale flow $k_U$. Representative examples of the flow patterns
realised in runs with small, medium, and high Reynolds numbers (from
left to right) and
forcing wavenumbers $\tkf = (3,5,10,30)$ (from top to bottom,
Sets~D-G) are shown
in
Fig.~\ref{fig:boxes}.
We also typically evolve the test-field equations in our runs so the
results pertaining to $\nut$ and $\etat$ are always obtained from the
same simulation.
All of our runs are listed in Tables~\ref{tab:runs1}--\ref{tab:runs3}.

In \Fig{fig:pakanut}
we show representative results for $a_2$ and $\nut$ obtained with the
methods M1 and M2a from Sets~A--C
(see Table~\ref{tab:runs1})
with forcing wavenumbers $3$, $5$,
and $10$.
The
coefficient $a_2$, corresponding to the AKA-effect, is consistent with
zero for all values of shear and with both methods that can detect
it. This conclusion applies to all of our models
and is consistent with the Galilean invariance of our forcing.

We note that,
as 
$a_2$ from method M2a is always very small, it has
a negligible effect on the quality of the fit and the value of $\nut$
in comparison to method M2b. For simplicity, we present results
obtained using M2b in what
follows. 
Overall,
no statistically significant values
were obtained for the coefficients $a_1$, $a_3$, $n_1$, and $n_3$.

\subsection{Turbulent viscosity}

\subsubsection{Dependence on position and sensitivity to shear}

The turbulent viscosities 
obtained for the two zeros employed in M1 (${\mbox M1}_1$ and ${\mbox M1}_2$ in \Fig{fig:pakanut}) agree
within error estimates and they agree with those obtained from M2a and M2b.
This suggests that $\nut$ has only a weak
dependence on $z$ or that the spatial profile of the turbulent
viscosity is such that it is not captured by this method. 

When the amplitude of the shear flow is varied,
the values of $\nut$ start to increase rapidly at the largest values
of $\Sh$; see \Fig{fig:pakanut}.
This is because the Navier--Stokes equations are inherently non-linear.
Therefore, imposing a
large-scale flow 
has an impact
on the turbulence. However, if the
shear is sufficiently weak, such feedback is small and reliable results
for $\nut$
can be obtained. To assess this question, we perform simulations at fixed
kinematic viscosity and given forcing wavenumber $\kf$ while varying the
shear systematically. With the other quantities held unchanged, the fluid
Reynolds number is a measure of the rms--velocity of the turbulence.
In \Fig{fig:pSh}, we
show the Reynolds numbers realised in the same sets as in
\Fig{fig:pakanut}.
We find that $\Rey$
increases mildly as a function of $\Sh$ for weak shear $(\Sh\lesssim 0.1)$
and starts to increase sharply at higher values
while the location of the transition
depends weakly on the forcing wavenumber such that the larger the
$\kf$, the smaller $\Sh$ is needed for the increase to occur.

The increase of $\Rey$ is due to the fact that the turbulence becomes
increasingly affected by the imposed shear and attains significant
anisotropy. In some cases with the highest values of $\Sh$, we also see
large-scale vorticity generation, which is likely related to what is known as
vorticity dynamo \cite[e.g.][]{KMB09}. Such hydrodynamic instability
can be excited by the off-diagonal components of the turbulent
viscosity tensor in anisotropic turbulence in the presence of
shear \citep{EKR03,EGKR07}.

These tests suggest that values of $\Sh$ sufficiently below 0.1 are
needed for the influence of the shear on the turbulence to remain
weak.
However, the excitation
condition of the vorticity dynamo
manifestly depends on the scale
separation ratio and likely also on the Reynolds number.
In our runs, we choose a constant value of $\Sh_{\rm c}$ for which $\Sh$
remains clearly below the excitation threshold.
Note that $\Sh_{\rm c}=\Sh\,\Ma$.
Another factor supposedly contributing at large Reynolds numbers is
shear-produced turbulence -- possibly through some sort of
finite amplitude instability.
Given that the shear strengths (in terms of $\Sh$)
considered here are relatively small, this effect is likely to
be weak in comparison to the turbulence production due to the applied
forcing.

\subsubsection{Dependence on $\Rey$}
Results for the turbulent viscosity as a function of the fluid
Reynolds number are shown in Fig.~\ref{fig:pnut_Re} for Sets~D--G
(see Table~\ref{tab:runs2}). Here the value of the shear parameter
$\Sh_{\rm c}$ is constant in each set. Additionally, the relaxation
time $\tau \urms \kf =1$ is kept
fixed by adjusting $\tau$, and $\tkf$ is varied between 3 (Set~D) and 30
(Set~G). Furthermore, these runs use
$\tkU=\tkB=1$.

We find that for low $\Rey$ and poor scale separation the signal is
noisy and produces large errors in $\nut$ unless very long time series
are produced. The runs with
$\kf\approx3$
and $\Rey\approx1$
were
in all sets
 typically run for several thousand turnover times whereas for
larger Reynolds numbers and scale separations the integration times
can be an order of magnitude shorter. The results
in the
low Reynolds
number regime are in agreement with $\nut \propto \Rey$ as expected
from
analytic studies using FOSA \citep{KR74a}. The
value of $\nut$
increases until $\Rey\approx 10$ after which it
saturates roughly to a constant
between one and two times $\nutz$ depending on the
scale separation ratio ${\mathscr S}$.
 However, we still see a slow
decrease
for the highest values of $\Rey$ which likely indicates
that even the highest resolution simulations are not in the regime of
fully developed turbulence. We note that the Mach
number changes by a factor between roughly two (Set~D) to four (Set~F)
between the extreme runs in each set. However, $\Ma$ saturates in the
high-$\Rey$ runs so compressibility effects are unlikely to explain
the slow
declining trend of $\nut$.

There is also a dependence on the scale separation ratio such that
higher values of 
${\mathscr S}$ result in larger values of $\tilde\nu_{\rm t}$. In
theory $\nut$ should converge towards the value at infinite scale
separation. This is confirmed by Sets~F and G where $\tkf=10$ and $30$,
respectively.

\subsubsection{Results from M3}

\begin{figure}[t]
\centering
\includegraphics[width=\columnwidth]{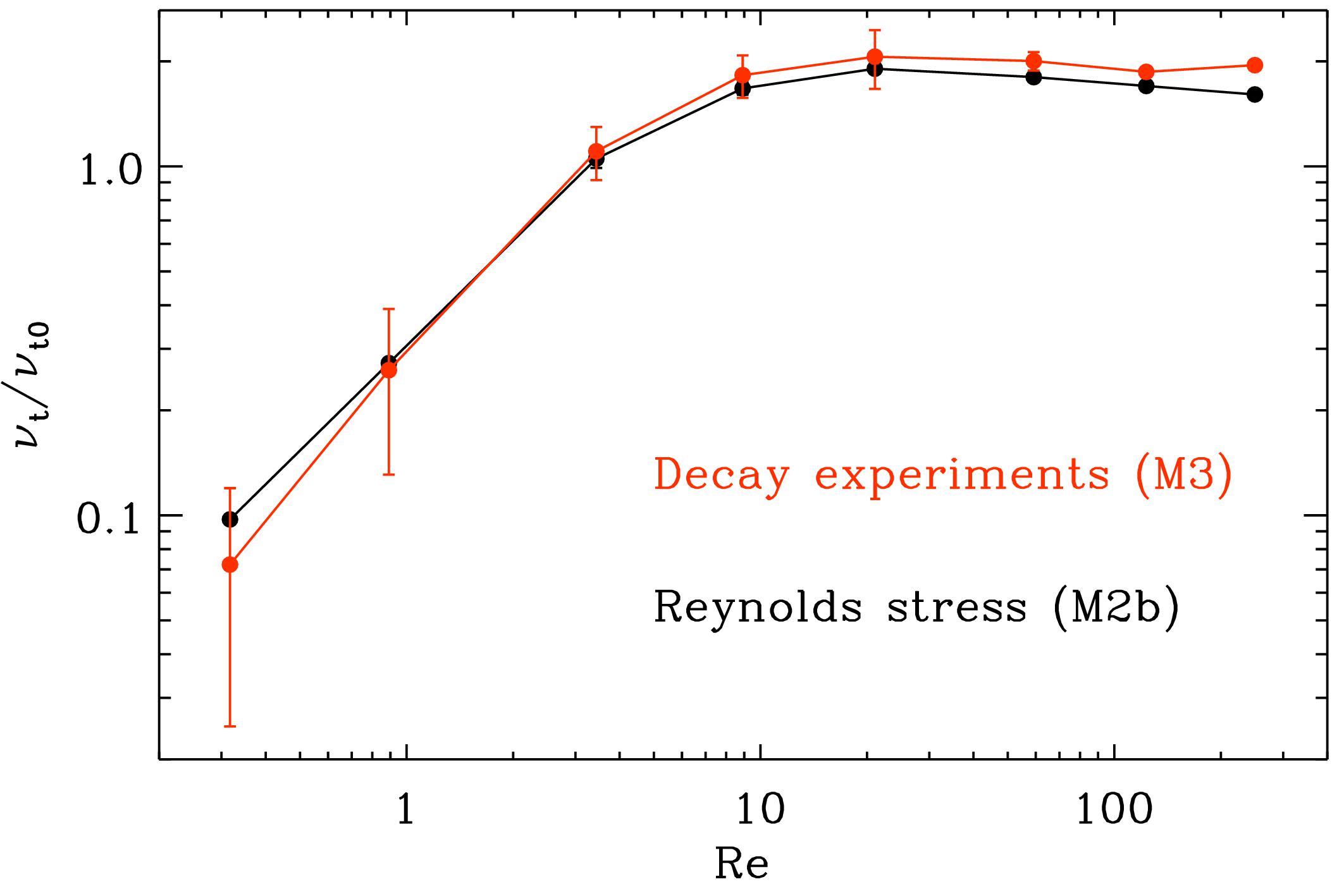}
\caption{Turbulent viscosity as a function of $\Rey$
from
 method M2b,
  Set~F (black line) and
  from corresponding decay experiments
  (method M3),
  Set~Fd (red).}
\label{fig:pnut_decay}
\end{figure}

We compare the results for
$\nut$
 from methods M2b and M3
in \Fig{fig:pnut_decay} for Sets~F and Fd. The
runs of the
latter were set up such
that $N=10$ snapshots from each of the runs in Set~F with imposed shear
were used as initial conditions. Thus each run in Set~F works as a
progenitor to ten decay experiments with the same system parameters in
Set~Fd.
We find that the results from methods M2b and M3 coincide within the
error estimates for low and intermediate Reynolds numbers
($\Rey\lesssim 20$). However, there is a systematic tendency for the
$\nut$ from the decay experiments to exceed the value from the
Reynolds stress method for $\Rey\gtrsim30$ by $10$--$20$ per cent.

\subsubsection{Dependence on scale separation ratio}

\begin{figure}[t]
\centering
\includegraphics[width=\columnwidth]{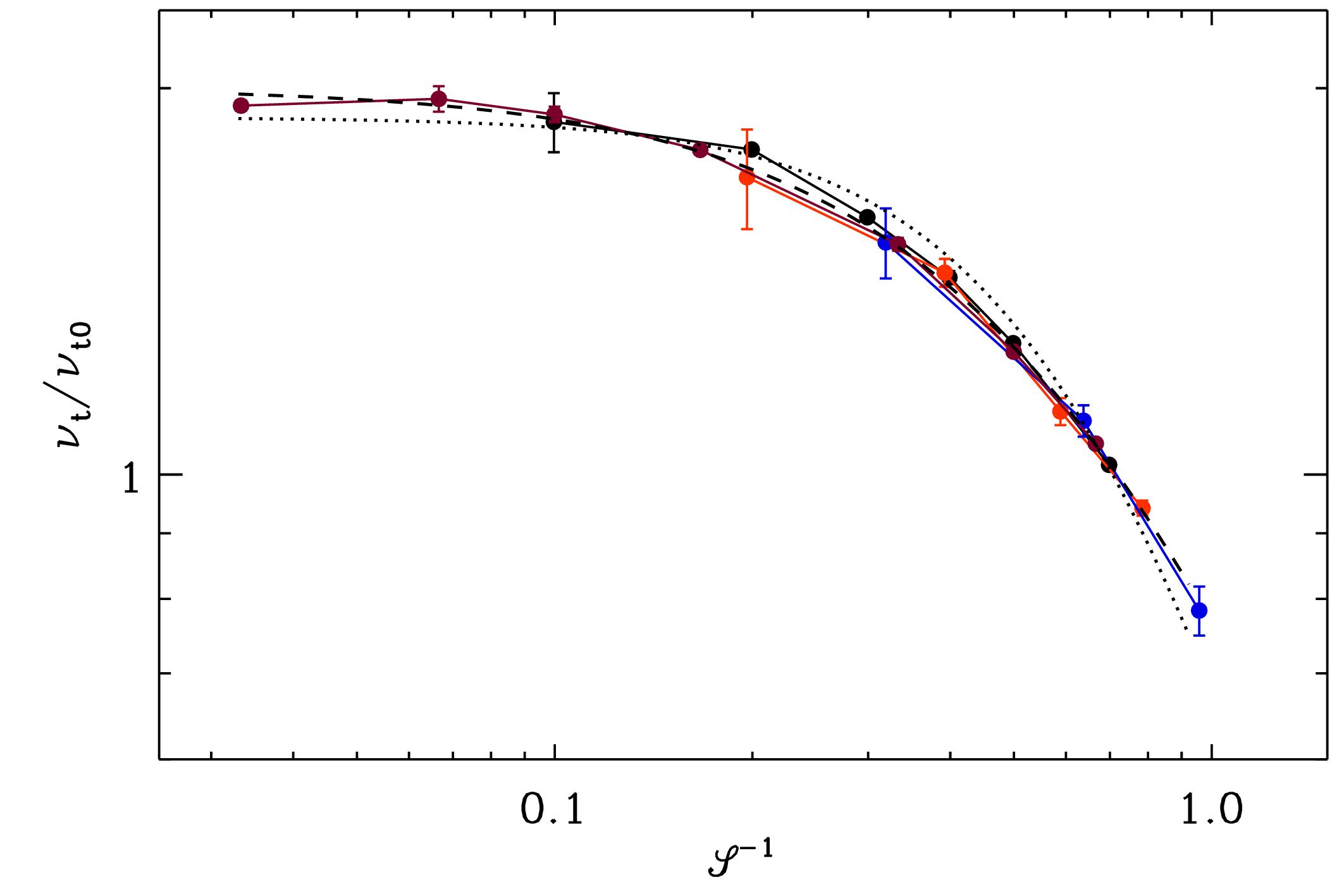}
\caption{Turbulent viscosity as a function of
  the 
  inverse
  scale separation ratio
   $k_U/\kf$
   for
   the four normalised forcing wavenumbers
   $\tkf=3$ (blue), 5 (red), 10 (black), and 30
   (purple), corresponding to Sets~H--K, respectively. Fits
     according to Eqs.~(\ref{equ:nutku}) and (\ref{equ:nutku2}) are
     shown by dotted and dashed lines, respectively.}
\label{fig:pnut_ku}
\end{figure}

The dependence of $\nut$ on the scale of the imposed velocity
$2\pi/k_U$
for
four
different forcing scales
is given
 in Table~\ref{tab:runs3}
(Sets~H--K).
We fit the data to a Lorentzian as a function of
the scale separation ratio ${\mathscr S}$
\begin{eqnarray}
  \nut(k_U/\kf) = \frac{\nut(0)}{1+\sigma(k_U/\kf)^2}
  = \frac{\nut(0)}{1+\sigma {\mathscr S}^{-2}},
\label{equ:nutku}
\end{eqnarray}
where $\nut(0)$ and $\sigma$ are fit
  parameters\footnote{The parameter $\sigma$ determines the
  intersection position of the asymptotics $\nut(0)$ and $\nut(0)
  {\mathscr S}^{2}/\sigma$ as 
  ${\mathscr S}=\sqrt{\sigma}$ 
  where $\nut$ becomes $\nut(0)/2$.}.
 We also fit the data to a more general function where
  the exponent $c$ of
   $k_U/\kf$
   is another fit parameter:
\begin{eqnarray}
\nut(k_U/\kf) = \frac{\nut(0)}{1+\sigma(k_U/\kf)^c}
= \frac{\nut(0)}{1+\sigma {\mathscr S}^{-c}}.
\label{equ:nutku2}
\end{eqnarray}
The inverse relative errors $\nut/\delta\nut$ are used as weights
  in the fit. The data and the fits are shown in
  \Fig{fig:pnut_ku}. While the data is in reasonable agreement with a
  Lorentzian with $\nut(0)=1.90$ and $\sigma=1.80$, the more general
  function with $\nut(0)=2.00$, $\sigma=1.61$, and $c=1.44$ yields a
  somewhat better fit.
Data consistent with Lorentzian behaviour has been found earlier for
$\etat$ in low Reynolds number turbulence \citep{BRS08}; see
\Table{tab:examples} for an overview of the $\sigma$ values found previously
in various cases ranging from magnetic diffusion in isotropic turbulence
to passive scalar diffusion in shear flows, in which $\sigma$ was typically
below unity.
However, a value of $c$ close to 4/3, as found here, is indeed
expected for fully developed Kolmogorov turbulence; see the
discussion in the conclusions of \cite{MB10}.

Note that $\sigma$ was found to be larger under the second-order
correlation approximation (SOCA), but the reason for this 
departure
is unclear.
Also, looking at \Table{tab:examples}, there is no obvious connection
between the values of $\sigma$ in different physical circumstances.
More examples 
are
needed to assess the robustness of the results
obtained so far.

Knowing the value of $\sigma$
is
important for
more accurate mean-field modelling.
In physical space, a prescription like \Eq{equ:nutku} corresponds
to a convolution, which makes the Reynolds stress at a given position 
dependent on the mean velocity within a certain neighbourhood.
In that way, non-locality is modelled. This is generally ignored in 
the common use of turbulent viscosity, although some attempts have been
made to include such affects to leading order \citep{1982ZaMM...62...95R}.
Ignoring non-locality corresponds to the limit $\sigma\to0$
or $k_U/\kf \to 0$,
which is often a questionable
assumption;
see \cite{BC18} for a discussion in
the context of spherical mean-field dynamos.

\begin{table}[t!]
\centering
\caption[]{Summary of the runs with varying scale of the shear,
$2\pi/k_U$.}
      \label{tab:runs3}
      \vspace{-0.5cm}
     $$
         \begin{array}{p{0.05\linewidth}ccrrcccccccc}
           \hline
           \noalign{\smallskip}
           Run & \Rey & \Sh & \Sh_{\rm c} [10^{-4}]\!\!\!\! & \tkf\!  & {\mathscr S}\!\! & \Ma & \tnut & \tetat & \Pmt & \mbox{Grid} \\ \hline
           H1 &  20 & 0.077 &  81 &  3  & 3.0 & 0.106 & 1.517 & 1.894 & 0.801 & 144^3 \\ 
           H2 &  20 & 0.153 & 163 &  3  & 1.5 & 0.106 & 1.101 & 1.477 & 0.746 & 144^3 \\ 
           H3 &  20 & 0.229 & 244 &  3  & 1.0 & 0.106 & 0.784 & 1.168 & 0.671 & 144^3 \\ 
           \hline
           I1 &  20 & 0.047 &  50 &  5  & 5.0 & 0.106 & 1.705 & 2.037 & 0.837 & 144^3 \\ 
           I2 &  20 & 0.094 & 100 &  5  & 2.5 & 0.107 & 1.436 & 1.782 & 0.806 & 144^3 \\ 
           I3 &  20 & 0.140 & 150 &  5  & 1.7 & 0.107 & 1.120 & 1.538 & 0.728 & 144^3 \\ 
           I4 &  20 & 0.187 & 200 &  5  & 1.3 & 0.107 & 0.942 & 1.337 & 0.704 & 144^3 \\ 
           \hline
           J1 &  21 & 0.024 &  25 &  10 &  10 & 0.106 & 1.883 & 2.142 & 0.879 & 144^3 \\ 
           J2 &  21 & 0.048 &  51 &  10 & 5.0 & 0.106 & 1.790 & 2.009 & 0.891 & 144^3 \\ 
           J3 &  21 & 0.072 &  76 &  10 & 3.3 & 0.106 & 1.584 & 1.899 & 0.834 & 144^3 \\ 
           J4 &  21 & 0.095 & 102 &  10 & 2.5 & 0.107 & 1.421 & 2.111 & 0.673 & 144^3 \\ 
           J5 &  21 & 0.119 & 127 &  10 & 2.0 & 0.107 & 1.265 & 1.635 & 0.774 & 144^3 \\ 
           J6 &  21 & 0.166 & 178 &  10 & 1.4 & 0.107 & 1.015 & 1.396 & 0.727 & 144^3 \\ 
           \hline
           K1 &  16 & 0.008 & 8.4 &  30 &  30 & 0.102 & 1.930 & 2.215 & 0.871 & 288^3 \\ 
           K2 &  17 & 0.016 &  17 &  30 &  15 & 0.103 & 4.343 & 6.481 & 0.670 & 288^3 \\ 
           K3 &  16 & 0.025 &  25 &  30 &  10 & 0.102 & 1.981 & 2.286 & 0.867 & 288^3 \\ 
           K4 &  16 & 0.042 &  42 &  30 & 6.0 & 0.102 & 1.790 & 2.096 & 0.854 & 288^3 \\ 
           K5 &  17 & 0.083 &  85 &  30 & 3.0 & 0.103 & 1.512 & 1.855 & 0.815 & 288^3 \\ 
           K6 &  17 & 0.123 & 127 &  30 & 2.0 & 0.103 & 1.247 & 1.628 & 0.766 & 288^3 \\ 
           K7 &  17 & 0.164 & 170 &  30 & 1.5 & 0.103 & 1.057 & 1.426 & 0.741 & 288^3 \\ 
           \hline
         \end{array}
     $$
         \tablefoot{All quantities have the same meanings as in Table~\ref{tab:runs1}.}
\end{table}

\begin{table}[t!]
\centering
\caption[]{Examples of $\sigma$ values found previously in other cases.}
      \label{tab:examples}
      \vspace{-0.5cm}
     $$
         \begin{array}{p{0.06\linewidth}lll}
           \hline
           \noalign{\smallskip}
$\sigma$ & \mbox{case}  & \mbox{Reference} \\[.5mm] \hline\\[-2.5mm]
0.16 & \mbox{shear flow, passive scalar} & \mbox{\cite{MB10}} \\
0.38 & \mbox{passive scalar} & \mbox{\cite{BSV09}} \\
0.25 & \mbox{Roberts flow, $\etat$, non-SOCA} & \mbox{\cite{BRS08}} \\
 1   & \mbox{Roberts flow, $\alpha$ \& $\etat$, SOCA} & \mbox{\cite{BRS08}} \\
1.80 & \mbox{$\nut$, isotropic} & \mbox{present work} \\
         \end{array}
     $$
\end{table}

\begin{figure}[t]
\centering
\includegraphics[width=\columnwidth]{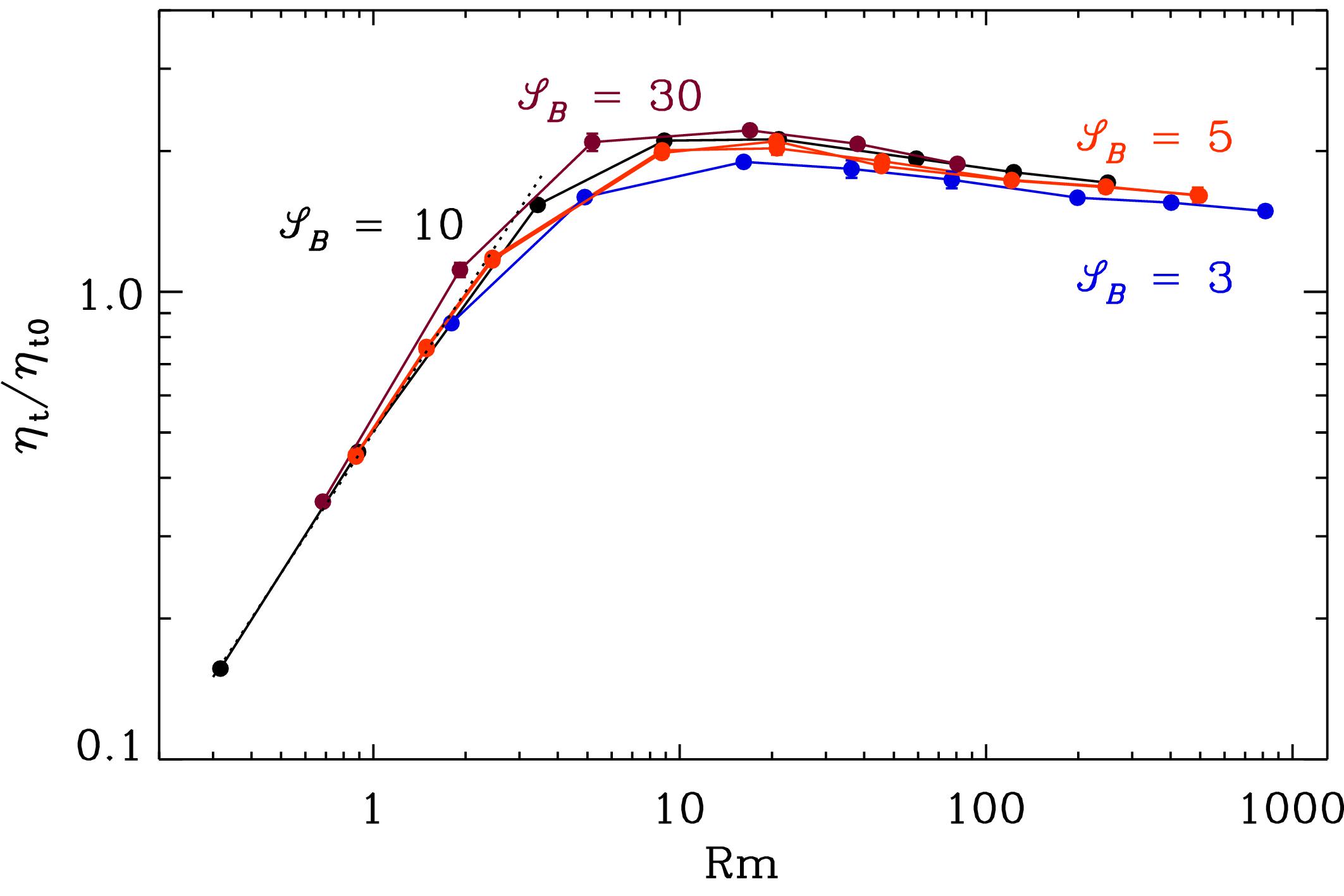}
\caption{Turbulent diffusivity $\etat$,  normalised by $\etatz$, as a function of the magnetic Reynolds
  number $\Rm$ for runs with $\Sh_{\rm c}=\,$
  const. within each of the Sets
  D (blue),
  E (red), F
  (black), and G (purple). The dotted black line is proportional to
  $\Rm$ and ${\mathscr S}_B=\kf/k_B$.}
\label{fig:petat_Re}
\end{figure}

\subsection{Turbulent diffusivity $\etat$}
\label{sec:tudi}

The turbulent magnetic diffusivity $\etat$ from Sets~D--G is shown in
Fig.~\ref{fig:petat_Re}. We find a similar qualitative behaviour as for
$\nut$ so that for small magnetic Reynolds numbers the value of
$\etat$ is proportional to $\Rm$ and the results converge when
the scale separation ratio is increased. As in the case of the
turbulent viscosity, we find a weak declining trend as a function of
$\Rm$
at its highest values
which was neither observed
by \cite{SBS08} in similar simulations without shear nor by
\cite{BRRK08} and \cite{MKTB09} in runs where the large-scale flow was
imposed via the shearing-box approximation. However, the error
estimates in the aforementioned studies are clearly greater than in
the present one and
thus a weak decreasing trend as a
function of $\Rm$
cannot be ruled out.
 Furthermore, the shear flows in the present
simulations are significantly weaker than in the cases of
\cite{BRRK08} and \cite{MKTB09}, such that their influence on the
turbulent transport coefficients is also weaker.

\begin{figure}[t]
\centering
\includegraphics[width=\columnwidth]{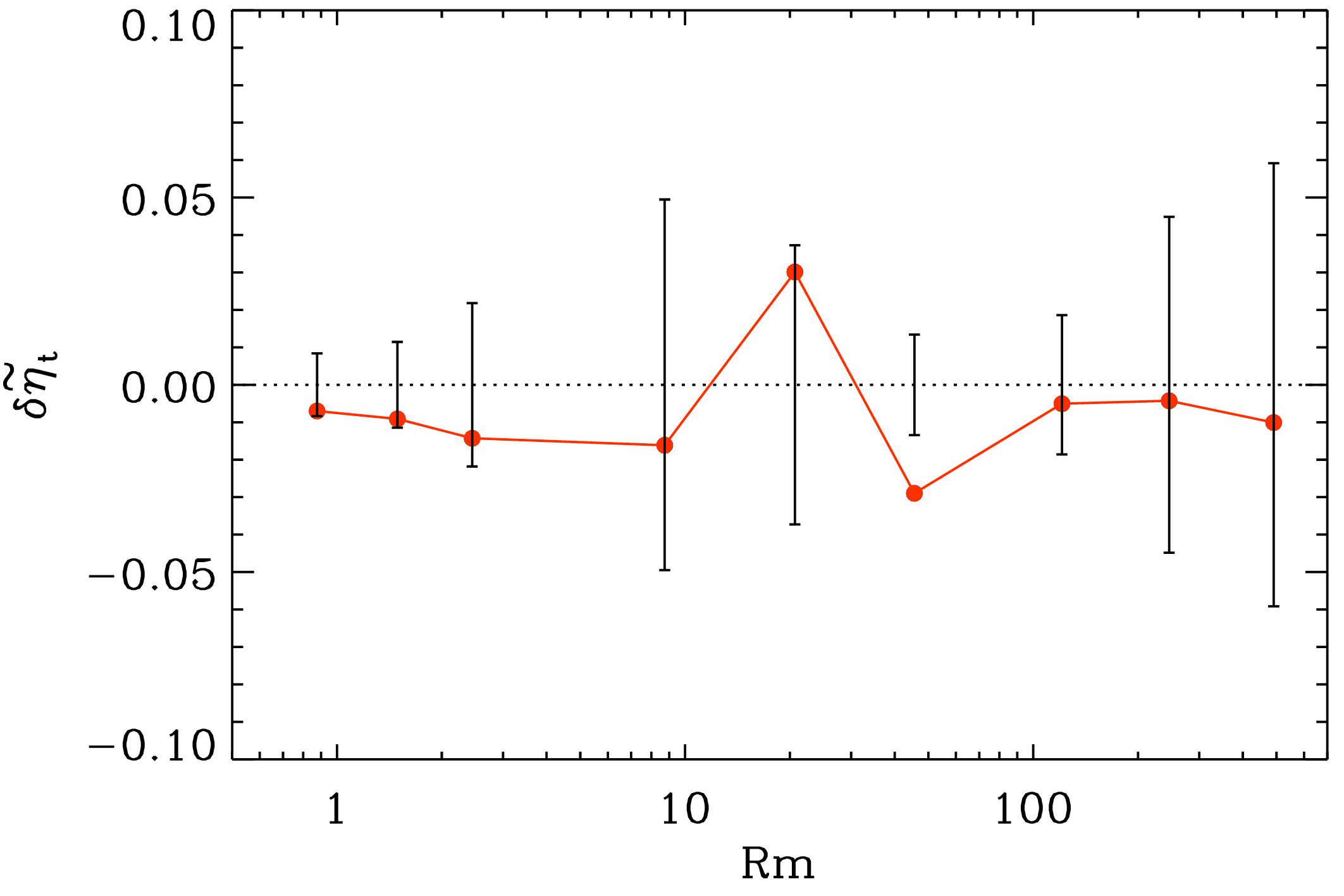}
\caption{
Relative
difference $\delta\tilde{\eta}_{\rm
    t}=(\etat^{(0)}-\etat)/\etat$ of turbulent magnetic diffusivity from
  Set~F and a corresponding set without shear, here denoted by
  superscript zero. The vertical bars indicate the error estimates of
  $\tilde{\eta}_{\rm t}$ from Set~F.}
\label{fig:petat_diff}
\end{figure}

We assess the effect of the shear
flow
 on the results by performing an
additional set of simulations
in which it
is omitted,
but otherwise the same
parameters as in Set~F
are employed.
We show the
results for the
difference of $\etat$ in these sets in Fig.~\ref{fig:petat_diff}. The
difference is typically of the order of a few per cent such that in most
cases the value from the case with shear is greater.
This is of the same order of magnitude as the error estimates for
$\etat$.
Thus we conclude that the systematic error due to the large-scale
anisotropy induced by the shear flow is insignificant in the
determination of the turbulent diffusivity.

\subsection{Turbulent magnetic Prandtl number}

\begin{figure}[t]
\centering
\includegraphics[width=\columnwidth]{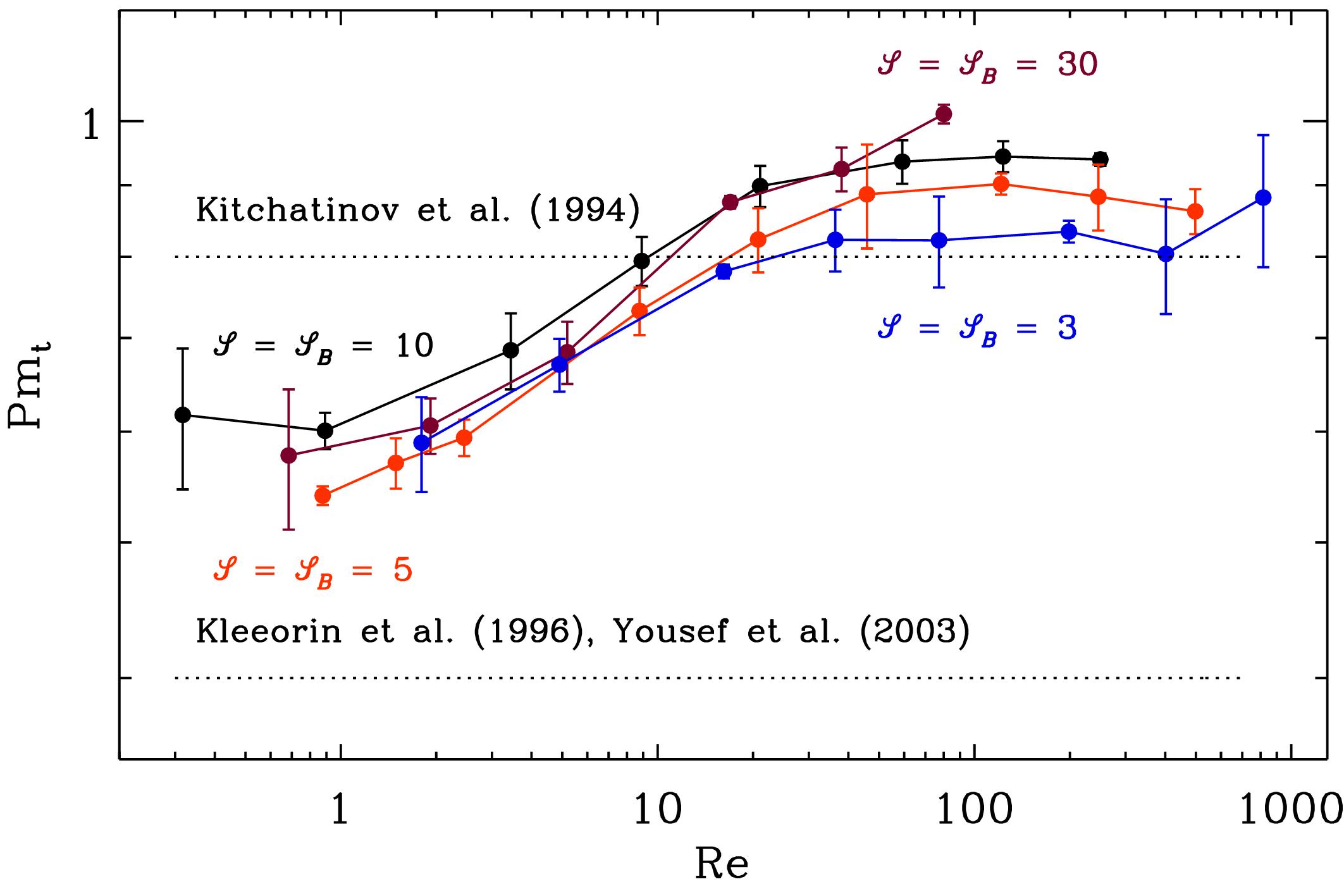}
\caption{Turbulent magnetic Prandtl number $\Pmt$ as a function of the
  Reynolds number $\Rey$ for the same sets of runs as in
  Fig.~\ref{fig:pnut_Re}. $\Pm=1$ is used in all runs. The dotted
  horizontal lines indicate the extrema of analytical results from
  different methods; see \Table{tab:Pmtexamples}.}
\label{fig:pPm_t_Re}
\end{figure}

Our results for $\Pmt$ as a function of Reynolds number and scale
separation ratio ${\mathscr S}$ 
are shown in Fig.~\ref{fig:pPm_t_Re}.  We
find that $\Pmt$ for $\Rey\gtrsim20$ is roughly a constant for each
value of
$\tkf$
while increasing
from roughly 0.8 for $\tkf=3$ to 0.95 for
$\tkf=10$.
Especially at low $\Rey$,
the
convergence
with respect to the scale
separation is
not as clear as
for $\nut$ and $\etat$ individually.
With respect to low Reynolds numbers,
we see an increasing trend starting from values between
0.55 and 0.65
at  $\Rey\approx 5$ until $\Rey\approx 20$.
At even lower
$\Rey$ the uncertainty in the
determination of $\nut$ becomes larger and the values of $\Pmt$ have
substantial error margins.

The turbulent magnetic Prandtl number has been computed with various
analytical techniques, see \Table{tab:Pmtexamples}. Considering the
limit of $\nu\rightarrow0$ or $\Rey\rightarrow\infty$, different
flavours of FOSA yield either $\Pmt=0.8$ \citep{KPR94} or $0.4$
\citep{YBR03}, whereas results MTA for fully turbulent convection
favours lower values \citep{2006GApFD.100..243R}. A similar spread of
values from $\Pmt\approx0.42$ \citep{2001PhPl....8.3945V} to
$\approx0.7-0.8$ \citep{FSP82,KR94,2011PhRvE..84d6311J} has been
reported using
renormalisation group methods for the case of three spatial dimensions
and weak magnetic fields.

Particularly at high scale separation,
our results are not compatible with any of the analytic results but
indicate a higher value than all of the theories.
This can be due to the fact that the turbulence in the
simulations is not in the fully developed regime and because the scale
separation achieved is still insufficient. Furthermore, analytic
theories must resort to approximations
that cannot be justified in high-Reynolds number turbulence.

\begin{table}[t!]
\centering
\caption[]{Comparison of values of $\Pmt$ from analytic and numerical studies.}
      \label{tab:Pmtexamples}
      \vspace{-0.5cm}
     $$
         \begin{array}{p{0.15\linewidth}lll}
           \hline 
           \noalign{\smallskip}
$\Pmt$ & \mbox{method}  & \mbox{Reference} \\ \hline
0.4  & \mbox{FOSA} & \mbox{\cite{YBR03}} \\
0.8  & \mbox{FOSA} & \mbox{\cite{KPR94}} \\
0.23--0.46  & \mbox{spectral MTA, conv.} & \mbox{\cite{2006GApFD.100..243R}} \\
$\approx0.72$ & \mbox{Renormalisation group} & \mbox{\cite{FSP82}} \\
$\approx0.79$ & \mbox{Renormalisation group} & \mbox{\cite{KR94}} \\
$\approx0.42$ & \mbox{Renormalisation group} & \mbox{\cite{2001PhPl....8.3945V}} \\
$\approx0.71$ & \mbox{Renormalisation group} & \mbox{\cite{2011PhRvE..84d6311J}} \\
$\approx1$    & \mbox{DNS, dec. MHD turb.} & \mbox{\cite{YBR03}} \\
$\approx0.9$  & \mbox{DNS, high $\Rey$} & \mbox{present work} \\
0.55--0.65  & \mbox{DNS, low $\Rey$} & \mbox{present work} \\
         \end{array}
     $$
\end{table}

\subsection{Dependence on forcing time scale}

\begin{figure}[t]
\centering
\includegraphics[width=\columnwidth]{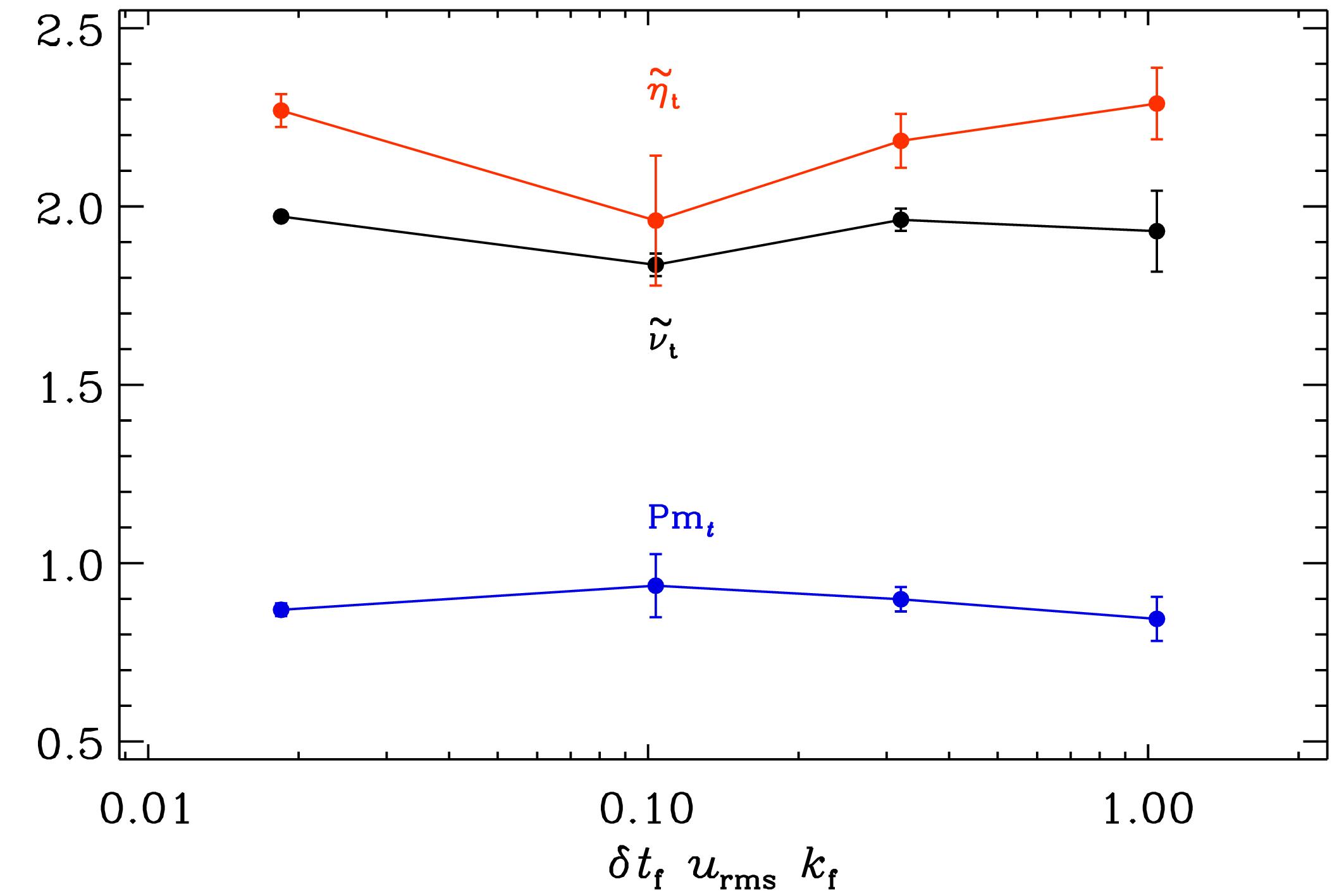}
\caption{Coefficients $\nut$ (black) and $\etat$ (red), normalised by
  $\nutz$ and $\etatz$, respectively; and $\Pmt$
    (blue) as functions of $\delta t_{\rm f} \kf \urms $ using
    Run~C4 as the fiducial model.}
\label{fig:pnut_dtforce}
\end{figure}

The bulk of the simulations considered here use a $\delta$
correlated random (white) forcing, see \Eq{equ:forcing}, such that a new
wavevector is chosen at every time step. It cannot, however, be
ruled out a priori that the results depend on the correlation
time of the forcing. Here we test the sensitivity of the results 
with respect to this correlation time by comparing the default case of
white--in--time forcing with cases where the forcing wavevector is
held constant for a time $\delta t_{\rm f}$. We take Run~C4 as
our fiducial model where $\delta t_{\rm f} \kf \urms\approx 0.02$
and  $\delta t_{\rm f}$
corresponds to the time step of the simulation. We increased $\delta
t_{\rm f} \kf \urms $ in steps by a factor of 50 altogether and computed
turbulent viscosity, magnetic diffusivity, and the 
magnetic Prandtl number; see \Fig{fig:pnut_dtforce}. We find that
$\nut$, $\etat$, and $\Pmt$ are essentially constant in this range
of parameters. 
Our method of switching the forcing with a period $\delta t_{\rm f}$ 
is crude because it induces discontinuities,
but the constancy of the coefficients suggests that robust results for the turbulent
transport coefficients are obtained nevertheless.

\section{Conclusions}
We have computed the turbulent viscosity ($\nut$) and magnetic
diffusivity ($\etat$) from simulations of forced turbulence using
imposed shear flows and the test-field method, respectively.
As expected, $\nut$ and $\etat$ are found to be proportional
to the respective Reynolds number at low $\Rey$ and $\Rm$.
With increasing values of $\Rey$ and $\Rm$,
the turbulent transport coefficients saturate at around
$\Rey \approx \Rm \approx 10$, but show a weakly decreasing trend
beyond.
The value of the turbulent viscosity estimated from the Reynolds stress, which
is interpreted to reflect the response of the system to a large-scale
flow, and from
the decay of a mean flow in the presence of turbulence
are in fair agreement. However, the
latter yields systematically slightly higher values for high Reynolds
numbers by less than 10 per cent.

The turbulent magnetic Prandtl number $\Pmt$ saturates between $0.8$
and $0.95$ for $\Rey\gtrsim10$ depending on the scale separation
ratio. We note that these
values are somewhat higher than those from the renormalisation group
approach and, especially, the first-order smoothing approach.
The value of $\Pmt$ computed here corresponds to the kinematic case
where the magnetic field is weak, which is often not the case in
astrophysical systems. Analytic studies predict quenching of turbulent
viscosity and magnetic diffusivity when the magnetic fields are
dynamically significant \citep[e.g.][]{KPR94}. The quenching of
$\etat$ has also been computed from numerical simulations
\citep[e.g.][]{KB09,2008ApJ...687L..49B,2014ApJ...795...16K}.
Similar studies for turbulent viscosity are so far lacking. Such
results will be reported elsewhere.

One of the other remaining issues to be addressed in the future
is the role
of compressibility effects,
in particular
that
of fluctuations of $\rho$.
In addition to making analytic progress by identifying potentially new
effects owing to
their presence,
it would be useful
to extend our simulations to the regime of larger Mach numbers.
Another possible extension of our work is to study the 
potential of a
 small-scale dynamo to give rise of magnetic stresses that could 
enhance the
turbulent viscosity, as has been suggested in the solar context to
alleviate the discrepancies between observations and simulations
of 
differential rotation and
convective velocities
\citep{2018arXiv180100560K}.

\begin{acknowledgements}
We thank the anonymous referee for the critical and constructive comments
on the manuscript.
The simulations were performed using the supercomputers hosted by CSC
-- IT Center for Science Ltd.\ in Espoo, Finland, who are administered
by the Finnish Ministry of Education. 
This work was supported by the
Deutsche Forschungsgemeinschaft Heisenberg programme (grant No.\ KA
4825/2-1; PJK), the Academy of Finland ReSoLVE Centre of Excellence
(grant No.\ 307411; PJK, MR, MJK), the NSF Astronomy and Astrophysics
Grants Program (grant 1615100), and the University of Colorado through
its support of the George Ellery Hale visiting faculty appointment.
This project has received funding from the European Research Council (ERC) 
under the European Union's Horizon 2020 research and innovation 
programme (project
 'UniSDyn',
 grant agreement No.\ 818665).
\end{acknowledgements}

\appendix
\section{Justification for \Eq{Qijdecomp}}
\label{app:deriv}

In the isothermal case, it is convenient to consider the quantity $H=\ln\rho$ instead of the density itself resulting in equations linear in $\overline H$ and free of triple correlations of
fluctuating quantities.
We denote the turbulence present under the condition $\mean{\bm U}=\nab \overline{H}=\boldsymbol{0}$ by $\uu^{(0)}$.
If $\uu^{(0)}$ describes isotropic turbulence, only the diagonal
components $Q_{ii}$ exist.
Proceeding to the situation $\mean{\bm U}= \boldsymbol{0}, \nab \overline{H}\ne\boldsymbol{0}$,
indicated by the superscript $(0,\mH)$
the fluctuations $\uu, h$ can be written as $\uu=\uu^{(0)}+\uu^{(0,\mH)}$,  $h=h^{(0)}+h^{(0,\mH)}$.
Correspondingly, the mean force is
\begin{alignat}{2}
    \mFFF &=&& \,2  \nu \,\overline{\boldsymbol{\mathsf s} \cdot \nab h} -  \overline{ \uu \cdot \nab \uu} = \mFFF^{(0)}  + \mFFF^{(0,\mH)}
\intertext{with $\boldsymbol{\mathsf s}=\boldsymbol{\mathsf S}-\mean{\boldsymbol{\mathsf S}}$, and}
    \mFFF^{(0)} &=&& \,2  \nu \,\overline{\boldsymbol{\mathsf s}^{(0)} \cdot \nab h^{(0)}} -  \overline{ \uu^{(0)} \cdot \nab \uu^{(0)}} , \\
    \mFFF^{(0,\mH)} &= && \,2 \nu \big( 
                    ( \overline{\boldsymbol{\mathsf s}^{(0)} + \boldsymbol{\mathsf s}^{(0,\mH)}) \cdot \nab h^{(0,\mH)} + \boldsymbol{\mathsf s}^{(0,\mH)} \cdot \nab h^{(0)}} \big) \nonumber \\
                    &&& -  \overline{ (\uu^{(0)}+\uu^{(0,\mH)}) \cdot \nab \uu^{(0,\mH)}} - \overline{\uu^{(0,\mH)} \cdot \nab \uu^{(0)}}.
\intertext{Restricting to first order in the mean quantities, we get}
   \mFFF^{(0,\mH)} &\,=&& \, 2\nu\big(
                    \overline{ \boldsymbol{\mathsf s}^{(0)} \cdot \nab h^{(0,\mH)} + \boldsymbol{\mathsf s}^{(0,\mH)} \cdot \nab h^{(0)}} \big) \nonumber\\
                    &&& -  \overline{ \uu^{(0)} \cdot \nab \uu^{(0,\mH)} } - \overline{ \uu^{(0,\mH)} \cdot \nab \uu^{(0)}}.
\end{alignat}
Analogously,  for $\mean{\bm U} \ne \boldsymbol{0}, \nab \overline{H} = \boldsymbol{0}$, we have
\begin{alignat}{2}
    \mFFF &=&& \mFFF^{(0)} + \mFFF^{(\mUUU,0)}
\intertext{with}
   \mFFF^{(\mUUU,0)} &\,=&& \, 
                    2\nu\left(\overline{ \boldsymbol{\mathsf s}^{(0)} \cdot \nab h^{(\mUUU,0)} + \boldsymbol{\mathsf s}^{(\mUUU,0)} \cdot \nab h^{(0)}} \right) \nonumber\\
                    &&& -  \overline{ \uu^{(0)} \cdot \nab \uu^{(\mUUU,0)}} - \overline{ \uu^{(\mUUU,0)} \cdot \nab \uu^{(0)}}
\end{alignat}
in first order.
If now both $\mean{\bm U}$ and $\nab \overline{H} $ do not vanish, one can see from the equations for the fluctuating quantities, again restricted to first order in the mean quantities,
\begin{alignat}{2}
  \partial_t h^{(\mUUU,\mH)} &= &&-\mUUU\cdot\nab h^{(0)} - \uu^{(0)}\cdot \nab \meanH -\nab\cdot  \uu^{(\mUUU,\mH)} \\
   &&& - \left( \uu^{(0)} \cdot \nab h^{(\mUUU,\mH)} + \uu^{(\mUUU,\mH)} \cdot \nab h^{(0)}\right)', \nonumber\\
  \partial_t \uu^{(\mUUU,\mH)} &= &&- \cs^2 \nab h^{(\mUUU,\mH)}  + \nu \nab^2 \uu^{(\mUUU,\mH)}  \\
   &&& + 2\nu \left( \boldsymbol{\mathsf s}^{(0)} \cdot \nab h^{(\mUUU,\mH)} + \boldsymbol{\mathsf s}^{(\mUUU,\mH)} \cdot \nab h^{(0)}\right)'\nonumber\\
   &&& -  \left( \uu^{(0)} \cdot \nab \uu^{(\mUUU,\mH)} + \uu^{(\mUUU,\mH)} \cdot \nab \uu^{(0)}\right)'  \nonumber \\
   &&& + 2\nu \,\left( \boldsymbol{\mathsf s}^{(0)}\cdot \nab \meanH +  \overline{ \boldsymbol{\mathsf S}}\cdot\nab h^{(0)} \right) \\
   &&& -   \uu^{(0)} \cdot \nab \mUUU - \mUUU \cdot \nab \uu^{(0)}, \nonumber
\intertext{that due to their linearity in $h^{(\mUUU,\mH)} $, $\uu^{(\mUUU,\mH)}$, and the additivity of their inhomogeneities containing $\mUUU$ and $\mH$, respectively, }
   h^{(\mUUU,\mH)} &= && \,h^{(0,\mH)} + h^{(\mUUU,0)}, \\
  \uu^{(\mUUU,\mH)} &= && \,\uu^{(0,\mH)} + \uu^{(\mUUU,0)},
\intertext{holds and hence}
  \mFFF &=&&\; \mFFF^{(0)} + \mFFF^{(\mUUU,0)} + \mFFF^{(0,\mH)}
\end{alignat}
to first order in $\mUUU$ and $\mH$.
Given that the major part of $ {\boldsymbol{\cal F}}$ is $\nab\cdot \boldsymbol{Q}$, an equivalent relationship can be assumed for $\boldsymbol{Q}$.

\bibliographystyle{aa}
\bibliography{bib}


\end{document}